\documentclass[12pt]{article}
\usepackage{amsmath}
\usepackage{amsfonts}
\usepackage[usenames]{color}
\usepackage{setspace}
\usepackage{multirow}
\usepackage{amssymb}

\usepackage{xfrac}

\usepackage{booktabs}
\usepackage[utf8]{inputenc}
\usepackage{lmodern}

\usepackage[font=small,skip=0pt]{caption}

\usepackage[backend=biber]{biblatex}
\usepackage{color, colortbl} % Highlighting the table rows/colomns
\usepackage[table]{xcolor}
\usepackage[first=0,last=9]{lcg}

\usepackage{graphicx}
\usepackage{caption}
\usepackage{subcaption}

\usepackage[section]{placeins}

\newtheorem{theorem}{Theorem}%[section]
\newtheorem{Definition}{Definition}

\newtheorem{Prop}{Proposition}

\newtheorem{lem}{Lemma}%[section]
\newtheorem{exmp}{Model}

\newcommand\independent{\protect\mathpalette{\protect\independenT}{\perp}}
\def\independenT#1#2{\mathrel{\rlap{$#1#2$}\mkern2mu{#1#2}}}

\usepackage{lipsum}
\makeatletter
\def\ps@pprintTitle{%
 \let\@oddhead\@empty
 \let\@evenhead\@empty
 \def\@oddfoot{}%
 \let\@evenfoot\@oddfoot}
\makeatother

\newcommand{\vect}{\textup{vec}}

\newcommand{\Cov}{\textup{Cov}}

\newcommand{\Var}{\textup{Var}}
\newcommand{\E}{\textup{E}}

\usepackage{float}

\usepackage{lipsum}
\makeatletter
\def\ps@pprintTitle{%
 \let\@oddhead\@empty
 \let\@evenhead\@empty
 \def\@oddfoot{}%
 \let\@evenfoot\@oddfoot}
\makeatother

\makeatletter

\newcommand*{\addFileDependency}[1]{% argument=file name and extension
\typeout{(#1)}% latexmk will find this if $recorder=0
\@addtofilelist{#1}
\IfFileExists{#1}{}{\typeout{No file #1.}}
}\makeatother

\usepackage[
top    = 2.54cm,
bottom = 2.59999cm,
left   = 2.54cm,
right  = 2.54cm
]{geometry}

\usepackage[symbol]{footmisc}

\begin{document}

\title{\bf Fourier Methods for  Sufficient Dimension Reduction in   Time Series}

\author{\Large S. Yaser Samadi$^a$  \footnote{Corresponding author, Email: ysamadi@siu.edu (S. Yaser Samadi)}  \  and  T. Priyan De Alwis$^b$  \\
$^a$  \small School of Mathematical and Statistical Sciences, Southern Illinois University  Carbondale, IL\\
$^b$\small Department of Mathematical Sciences, Worcester Polytechnic Institute, 
Worcester, MA
}

\date{ }
\maketitle
%\date{ }

\setlength{\abovedisplayskip}{6.pt}
\setlength{\belowdisplayskip}{6.pt}

\begin{abstract}
Dimensionality reduction has always been one of the most significant and challenging problems in the analysis of high-dimensional data. In the context of time series analysis,  our focus is on the estimation and inference of conditional mean and variance functions. 
By using central mean and variance dimension reduction subspaces that preserve sufficient information about the response, one can effectively estimate the unknown mean and variance functions of the time series. While the literature presents several approaches  
to estimate the time series central mean and variance subspaces  (TS-CMS and TS-CVS), 
  these methods tend to be computationally intensive and infeasible for practical applications. % not feasible in practice.
  By employing the Fourier transform,  we derive explicit estimators for TS-CMS and TS-CVS. 
  These proposed estimators are demonstrated to be consistent,  asymptotically normal,  and efficient.  Simulation studies have been conducted to evaluate the performance of the proposed method. The results show that our method is significantly more accurate and computationally efficient than existing methods. Furthermore, the method has been applied to the Canadian Lynx dataset.
\end{abstract}

%\begin{keyword}

% \maketitle
\date{ }
%\end{frontmatter}

%\linenumbers
%\linenumbers
\section{\large  Introduction}
%\vspace{-.1in}
Complicated time-series data are ubiquitous in many applications ranging from economics and finance to meteorology,  medicine,  genetics,  and others.  The main goal of time series analysis is to make inferences about the conditional mean and variance functions of the current value given its past values.    Autoregressive moving average (ARMA) models (Brockwell and Davis,  1991) are the well-known classical linear models for analyzing time series data.  However,  there is an abundance of empirical evidence indicating that some complex features such as nonlinearity,  heteroscedasticity,  and asymmetry in many real time series data cannot be captured by the linear time series models,   resulting in the development of different types of nonlinear time series models.    Among them,  some have focused on parametric nonlinear time series models,  e.g.,  Engle (1982); Bollerslev (1986); Tong (1990); Granger and Ter{\"a}svirta (1993);   Tj{\o}stheim (1994); Tiao and Tsay (1994);  and some on nonparametric and semiparametric time series models,  e.g.,   Tong (1995); Masry and Fan (1997);  Xia and Li (1999); Fan and Yao (2003); and Gao (2007).
In order to circumvent the course of dimensionality,   Xia and Li (1999), and Xia et al. (1999, 2002) proposed a single-index model that addresses the issue of dimension reduction in time series.

Sufficient dimension reduction (SDR)  is a powerful tool for reducing dimensionality without losing information,  which was initially introduced for regression problems by Li (1991), then the SDR theory established by Cook (1994, 1998).   
In SDR studies, several estimation methods have been developed to estimate the central subspace (CS), the central mean subspace (CMS), and the central variance subspace (CVS) in regression problems. Initially, the sliced inverse regression (SIR) was introduced by Li (1991) to estimate the CS in regression.   There are various  methods for SDR in regression context  to mitigate the curse of dimensionality  including principal Hessian direction (pHd) (Li,  1992; Cook, 1998);   sliced average variance estimation (SAVE) (Cook and Weisberg, 1991); iterative Hessian transformation (IHT) procedure (Cook and Li,  2002);   minimum average variance estimation (MAVE)  (Xia et al.,  2002);   method of direction estimation (MODE) (Yin and Cook, 2005; Yin et al., 2008);   
  Fourier transformation (FT) method (Zhu and Zeng, 2006;  Weng and Yin,  2018); squared residuals-based outer product gradient (OPG) method to estimate CVS (Zhu and Zhu,  2009); integral transformation (ITM) procedure (Zeng and Zhu, 2010).

Although there are a variety of methods available in the literature for estimating the CS,  the CMS, and the CVS  in regression, there are a few approaches proposed in the literature to estimate these subspaces in time series, i.e.,   to estimate the time series  CS   (TS-CS), the time series CMS   (TS-CMS), and the time series CVS     (TS-CVS).   
Recently,  the theory of SDR has been developed for time series and some nonparametric methods have been proposed in the literature,   however,  these approaches are computationally intensive and expensive. 
 Park et al. (2010) extended the idea of the expected conditional loglikelihood approach of Yin and Cook (2005) to estimate the TS-CS by maximizing the Kullback-Leibler distance function.  Park et al.  (2009), used the Nadaraya-Watson (NW) kernel estimator for estimating the TS-CMS, and proposed a modified Schwarz Bayesian criterion (MSBC) to estimate both the lag order and the dimension of the TS-CMS.
Throughout the paper, we use the abbreviation NW to refer to this procedure.   The NW approach estimates the direction of the mean function by minimizing the residual sum of the squares function. Moreover, Park and Samadi (2014) proposed a nonparametric kernel method to estimate the TS-CVS.  
 They were motivated by the conditional variance function model in regression proposed by Zhu and Zhu (2009)  and used a similar method as the NW approach to estimate the TS-CVS by minimizing the sum of squared innovations. 
 Park and Samadi (2020) proposed a two-step procedure that considers the problem of making inferences about both the conditional mean and the conditional variance functions. Their two-step dimension reduction approach provides a different dimension reduction on the mean and variance functions.  
Because of the intensive computational requirements, however, these methods become impractical.

In this paper,  we propose a new fast and efficient algorithm for estimating the TS-CMS and  TS-CVS. We achieve this by utilizing the Fourier transformed gradient operator of the conditional mean and variance functions of time series.  Our proposed procedure does not require any specific model assumptions. 
  By using the Fourier transformation,  we are able to fully extract the TS-CMS and TS-CVS  without estimating the directions of the conditional mean and variance functions or their link functions. The estimation algorithm is constructed by obtaining a candidate matrix called $\mathbf{M}_{{FMT}}$ ($\mathbf{M}_{{FVT}}$) by imposing certain assumptions. Subsequently, the subspace spanned by the columns of the candidate matrix $\mathbf{M}_{{FMT}}$ ($\mathbf{M}_{{FVT}}$) is employed for estimating the TS-CMS (TS-CVS).  %($\mathcal{S}_{\E[y_t\vert\mathbf{Y}_{t-1}]}$).

Suppose  $\{y_t\}_{t=1}^N$ is an univariate time series process and let $\mathbf{Y}_{t-1}$ is the vector of lag variables, i.e., $\mathbf{Y}_{t-1}=(y_{t-1},\dots,y_{t-p})^T$. The main goal of time series analysis is to make inferences about the conditional distribution of $y_t$ given $\mathbf{Y}_{t-1}$ (that is, $y_t\vert \mathbf{Y}_{t-1}$), and the conditional mean and/or variance of $y_t$ given $\mathbf{Y}_{t-1}$ (that is,  $\E[{y_t\vert \mathbf{Y}_{t-1}}]$ and $\Var(y_t\vert \mathbf{Y}_{t-1})$,  respectively). 
% for reducing data dimensionality without stringent model assumptions.
The  idea of SDR is to find a $p\times d$ matrix $\boldsymbol\eta_d$, $d\leq p$, such that $p$-dimensional predictor vector $\mathbf{Y}_{t-1}$ can be replaced by the $d$-dimensional vector $\boldsymbol{\eta}_d^T\mathbf{Y}_{t-1}$ without loss of information. 
That is, the $d$-dimensional vector $\boldsymbol{\eta}_d^T\mathbf{Y}_{t-1}$ can be used instead of $p$-dimensional vector $\mathbf{Y}_{t-1}$  to make inferences about the response $y_t$ while preserving all information on response,  and if $d<p$, dimension reduction is achieved.   This can be explained by $\E[{y_t\vert \mathbf{Y}_{t-1}}]=\E[{y_t\vert \boldsymbol{\eta}_d^T\mathbf{Y}_{t-1} }]$. % and  $\Var(y_t\vert \mathbf{Y}_{t-1})=\Var(y_t\vert \boldsymbol{\eta}_d^T\mathbf{Y}_{t-1})$.
Moreover, the SDR technique targets reducing data dimensionality without requiring stringent model assumptions while preserving useful information contained in data (Cook,  2018). 
 
 The rest of the paper is organized as follows.  In Section 2,  we present preliminaries and definitions of CMS and CVS in time series.   In Section 3, we introduce notations for the Fourier transformation method in time series, where the majority of the theoretical framework and estimation procedure are expounded.      Section 4 delves into the discussion of the asymptotic properties of the proposed estimators. In Section 5,  new estimation algorithms are introduced to estimate the unknown lag order ($p$), the dimension of the minimal dimension reduction subspace ($d$), the tuning parameter, and finally the TS-CMS. The estimation process is demonstrated using %\color{black} three \color{black} 
 simulation examples and their results are presented and evaluated in Section 6. An empirical application is presented in Section 7.    All essential proofs are included in the Appendix, while some lengthy proofs and additional simulation results can be found in the Supplementary Material.

\section{ \large  Central Mean and Variance Subspaces in Time Series }

In a more general setting, we consider the time series model as  follows 
\begin{equation}\label{eq:4}
  y_t=g_1(\boldsymbol\eta_d^T\mathbf{Y}_{t-1})+x_t,
\end{equation}
where $g_1(\cdot)$ is an unknown  smooth link function, $\mathbf{Y}_{t-1}=(y_{t-1},\dots,y_{t-p})^T$, and $x_t$'s are white noise error terms,  or   $x_{t}$ has a general  heteroscedastic structure as
\begin{equation}\label{eq:5}
  x_t=\sqrt{g_2(\boldsymbol\Gamma_{d^{'}}^T\mathbf{X}_{t-1})}~~\varepsilon_t,
\end{equation}
with  $x_t=y_t-\E[y_t\vert \mathbf{Y}_{t-1}]$,   
where $\varepsilon_t$'s are white noise terms, $g_2(\cdot)$ is an unknown  smooth link function, and $\mathbf{X}_{t-1}=(x_{t-1}^2,\dots,x_{t-q}^2)^T$.

The central mean subspace for the conditional mean function is defined as a natural inferential and estimative object for dimensional reduction when the mean function $\E[y_t\vert \mathbf{Y}_{t-1}]$ is of interest. The CMS was introduced by Cook and Li (2002) in the regression context. 
 
 \begin{Definition}\label{def:2}
  Let $\mathcal{S}$ denote a subspace of $\mathbb{R}^p$ with projection $\mathcal{P_S}$,  and
let $\mathcal{P_S}: \mathbb{R}^p\to \mathcal{S}$   be the  orthogonal projection operator onto a $d\ (<p)$-dimensional subspace  $\mathcal{S}$. Then, $\mathcal{P_S}$ is a mean dimension reduction subspace (DRS) for the conditional mean if
$y_t \independent \E[y_t\vert \mathbf{Y}_{t-1}] \vert \mathcal{P_S}\mathbf{Y}_{t-1}
$, 
%\begin{equation}\label{eq:2}
%y_t \independent \E[y_t\vert \mathbf{Y}_{t-1}] \vert \mathcal{P_S}\mathbf{Y}_{t-1},
%\end{equation}
where $\independent$ indicates independence.  The intersection of all mean DRSs for time series is called the time series central mean DRS,  or in brief ``time series central mean subspace (TS-CMS)'', and is denoted by $\mathcal{S}_{\E[y_t\vert \mathbf{Y}_{t-1}]}$ with $d=\dim(\mathcal{S}_{\E[y_t\vert \mathbf{Y}_{t-1}]}) $.
\end{Definition}
Thus,  $y_t$ is independent of  $\E[y_t\vert \mathbf{Y}_{t-1}]$  given any value for  $\mathcal{P_S}\mathbf{Y}_{t-1}$. 
 The definition of TS-CMS can be extended to encompass higher conditional moments and in particular to the conditional variance. That is, the CVS for time series is characterized as a step in the inferential process of dimension reduction when the variance function of time series is of interest. Yin and Cook (2002) introduced the central $r$th moment subspaces in the regression context. Park and Samadi (2014) proposed the CVS for time series by considering the conditional mean function of the squared innovation time series $\{x^2_t\}_{t=1}^N$ defined in   \eqref{eq:5}.  % i.e., $\E[x_t^2 \vert \mathbf{X}_{t-1}]$, where $x_t=y_t-\E[y_t \vert \mathbf{Y}_{t-1}]$,  and $\mathbf{X}_{t-1}=(x^2_{t-1},\dots,x^2_{t-q})^T$.

\begin{Definition}\label{def:3}
Let $x_t=y_t-\E[y_t \vert \mathbf{Y}_{t-1}]$.
Then, an orthogonal projection $\mathcal{P_{\mathbf{S}'}}: \mathbb{R}^q\to \mathcal{\mathbf{S}'}$ onto a $d^{'}(<q)$-dimensional subspace  ${\mathcal{S}'}$ is a sufficient variance DRS if
$x_t^2 \independent \E[x_t^2\vert \mathbf{X}_{t-1}] \vert \mathcal{P_{S'}}\mathbf{X}_{t-1}$.
%\begin{equation}\label{eq:3}
%x_t^2 \independent \E[x_t^2\vert \mathbf{X}_{t-1}] \vert \mathcal{P_{S'}}\mathbf{X}_{t-1},
%\end{equation}
 The intersection of all variance  DRSs for time series is called the time series central variance DRS,  or in brief ``time series central variance subspace (TS-CVS)'', which is denoted by $\mathcal{S}_{\E(x_t^2\vert \mathbf{X}_{t-1})}$,  with $d^{'}=\dim(\mathcal{S}_{\E(x_t^2\vert \mathbf{X}_{t-1})})$.
\end{Definition}

 \begin{Prop}\label{prop:1.1}
 The following three conditions are equivalent. 
\begin{itemize}
  \item[(i)] $y_t \independent \E[y_t\vert \mathbf{Y}_{t-1}]\vert \boldsymbol\eta_d^T\mathbf{Y}_{t-1}$.  ~~~~~~~~~
   (ii) $\Cov(y_t,\E[y_t\vert \mathbf{Y}_{t-1}]\vert \boldsymbol\eta_d^T\mathbf{Y}_{t-1})=0$.
  \item[(iii)] $\E[y_t\vert \mathbf{Y}_{t-1}]$ is a measurable function of $\boldsymbol\eta_d^T\mathbf{Y}_{t-1}$.
\end{itemize}
\end{Prop}
The proof of   Proposition  \ref{prop:1.1}   is similar  to that   of Proposition 1 in Park et al. (2009), and hence is omitted.  
Similar results hold for the variance dimension reduction subspaces as follows.

\begin{Prop}\label{prop:2}
  The following three conditions are equivalent.\label{prop:1.2}
\begin{itemize}
  \item[(i)] $x_t^2 \independent \E[x_t^2\vert \mathbf{X}_{t-1}]\vert \boldsymbol\Gamma_{d'}^T\mathbf{X}_{t-1}$, $\left(\boldsymbol\Gamma_{d'} \in \mathbb{R}^{q\times d'}\right)$. ~~~~~
  (ii) $\Cov(x_t^2,\E[x^2_t\vert \mathbf{X}_{t-1}]\vert \boldsymbol\Gamma_{d'}^T\mathbf{X}_{t-1})=0$.
  \item[(iii)] $\E[x_t^2\vert \mathbf{X}_{t-1}]$ is a measurable function of $\boldsymbol\Gamma_{d'}^T\mathbf{X}_{t-1}$.
\end{itemize}
\end{Prop}

%%%%%%%%%%%%%%%%%%%%%%%%%%%%%%%%%%%%%%%%%%%%%%%%%%%%%%%%%%%%%%%%%%%%%%%%%%%%%%%%%%%%%%%%%%%%%%%%%%%%%%%%%%%%%
%%%%%%%%%%%%%%%%%%%%%%%%%%%%%%%%%%%%%%%%%%%%%%%%%%%%%%%%%%%%%%%%%%%%%%%%%%%%%%%%%%%%%%%%%%%%%%%%%%%%%%%%%%%%%%%%%
\section{\large  Fourier Method to Estimate TS-CMS and TS-CVS }\label{sec:3}

\setlength{\abovedisplayskip}{6.pt}
\setlength{\belowdisplayskip}{6.pt}
%\section{Fourier Method to Estimate Time Series Central Mean Subspace (TS-CMS) }\label{sec:3}
In this section, we focus on the problem of estimating the basis matrix $\boldsymbol\eta \in \mathbb{R}^{p \times d}$ of the  TS-CMS,  $\boldsymbol\eta=(\boldsymbol\eta_1,\dots,\boldsymbol\eta_d)$, by assuming that both the number of lags $p$ and the dimension of subspace, $d$, are known. We propose and illustrate a new estimation procedure to estimate $p$ and $d$ in Section \ref{Sec:5:1}.  Here, we provide the main conditions and assumptions required to derive a candidate matrix called $\mathbf{M}_{{FMT}}$, whose column space is equal to the TS-CMS.   In the following subsections, the Fourier transform-based estimation method of the  TS-CMS  is described.  Similar steps can be extended to construct a candidate matrix ($\mathbf{M}_{{FVT}}$) for estimating the TS-CVS (see Section \ref{sec:3:2}); however,  for brevity, we only describe the procedure of the TS-CMS in Section \ref{sec:3:1}. 
All necessary proofs of the Propositions and Theorems given in this section are presented in the Appendix and Supplement A.  %\ref{Sup:C}.
\subsection{\bfseries  Estimation of the Central Mean Subspace in Time Series}\label{sec:3:1}
Let $m(\mathbf{y}_{t-1}$)=$\E[y_t\vert \mathbf{Y}_{t-1}=\mathbf{y}_{t-1}]$ denotes the conditional mean function of the model in   \eqref{eq:4}. Suppose $\boldsymbol\eta=(\boldsymbol\eta_1,\dots,\boldsymbol\eta_d)$ is a basis of the TS-CMS  $\mathcal{S}_{\E[y_t\vert \mathbf{Y}_t]}$, where $\boldsymbol\eta$ is a $p\times d$ ($d\leq p$) dimensional matrix. Then, from   (\ref{eq:4}) we have
\begin{equation}\label{eq:6}
m(\mathbf{y}_{t-1})=\E[y_t\vert \mathbf{Y}_{t-1}=\mathbf{y}_{t-1}]=g_1(\boldsymbol\eta_1^T\mathbf{y}_{t-1},\dots,\boldsymbol\eta_d^T\mathbf{y}_{t-1}),
\end{equation}
where $g_1(\cdot)$ is a link function from $\mathbb{R}^d \text{ to } \mathbb{R}$. Let $\mathbf{u}_{t-1}=\boldsymbol\eta^T \mathbf{y}_{t-1}$, then $m(\mathbf{y}_{t-1})=g_1(\mathbf{u}_{t-1})$. Define the gradient operator as $\frac{\partial}{\partial \mathbf{y}_{t-1}}=(\frac{\partial}{\partial y_{t-1}},\dots,\frac{\partial}{\partial y_{t-p}})^T$, then by the chain rule of differentiation we have,
\begin{equation}\label{eq:7}
\frac{\partial}{\partial \mathbf{y}_{t-1}}m(\mathbf{y}_{t-1})= \boldsymbol\eta \frac{\partial}{\partial \mathbf{u}_{t-1}} g_1(\mathbf{u}_{t-1}).
\end{equation}
Thus, the gradient of $m(\mathbf{y}_{t-1})$ for given $\mathbf{y}_{t-1}$ can be written as $\frac{\partial}{\partial \mathbf{y}_{t-1}}m(\mathbf{y}_{t-1})=\sum_{k=1}^{d}c_k \boldsymbol\eta_k$ for some constants $c_1,\dots,c_d$. That is, gradient of $m(\mathbf{y}_{t-1})$ is a linear combination of the columns of $\boldsymbol\eta=(\boldsymbol\eta_1,\dots,\boldsymbol\eta_d)$, therefore $\frac{\partial}{\partial \mathbf{y}_{t-1}}m(\mathbf{y}_{t-1}) \in \mathcal{S}_{\E[y_t\vert \mathbf{Y}_{t-1}]}$.
\begin{theorem}\label{th:1}
	Let Supp$(\mathbf{Y}_{t-1})=\{ \mathbf{y}_{t-1} \in  \mathbb{R}^p:  f(\mathbf{y}_{t-1})>0\}$ be the support of $\mathbf{Y}_{t-1}$.  Then, both the set of all gradients of $m(\mathbf{y}_{t-1})$ with respect to $\mathbf{y}_{t-1} \in Supp (\mathbf{Y}_{t-1})$, and the set of all gradients of $m(\mathbf{y}_{t-1})$ weighted by $f(\mathbf{y}_{t-1})$,  span the TS-CMS ($\mathcal{S}_{\E[y_t\vert \mathbf{Y}_{t-1}]}$), that is
	\small
\begin{align}\label{eq:8}
\begin{split}
		\mathcal{S}_{\E[y_t\vert \mathbf{Y}_{t-1}]}&= \textup{span}\left\{ \frac{ \partial}{\partial \mathbf{y}_{t-1}}m(\mathbf{y}_{t-1}),~~\mathbf{y}_{t-1}\in Supp(\mathbf{Y}_{t-1})\right\}\\		
		 &=\textup{span}\left\{ f(\mathbf{y}_{t-1})\left( \frac{\partial}{\partial \mathbf{y}_{t-1}}m(\mathbf{y}_{t-1})\right),~~\mathbf{y}_{t-1} \in \mathbb{R}^p\right\}.
\end{split}
\end{align}
\normalsize
\end{theorem}
According to Theorem \ref{th:1}, we can estimate the TS-CMS  by estimating the gradient of the mean function $m(\mathbf{y}_{t-1})$ weighted by $f(\mathbf{y}_{t-1})$. However, this would not be an efficient way to estimate the TS-CMS, because this would require estimating the link function $g_1(\cdot)$ in   (\ref{eq:6}) and its derivatives nonparametrically that demands extra complex calculations to find the basis vectors of the subspace spanned by the gradients of the conditional expectation. Park et al. (2009) used  the  Nadaraya-Watson (NW) kernel estimator   %Kullback-Leibler divergence 
to estimate the directions of the time series mean function nonparametrically. Their method requires using iterative nonlinear optimization techniques to estimate the TS-CMS. The main drawbacks of this procedure are the low efficiency of the estimation process and high computational cost. Notice that, our goal is to extract the basis vectors of the TS-CMS and achieve dimension reduction with the least effort and without estimating the link function $g_1(\cdot)$ and its derivatives. Therefore, we use the Fourier transform algorithm to achieve our goal without doing unnecessary and computationally expensive calculations.  In the process of estimating the TS-CMS, the Fourier transform of the gradient of $m(\mathbf{y}_{t-1})$  is used instead of using $\frac{\partial}{\partial \mathbf{y}_{t-1}}m(\mathbf{y}_{t-1})$ directly. % The following paragraphs illustrate the process of using this approach to produce the candidate matrix $\mathbf{M}_{FMT}$.\\

Let $\boldsymbol\psi_t(\boldsymbol\omega)$ is the Fourier transform of the density-weighted gradient of $m(\mathbf{y}_{t-1})$ given as
\begin{equation}\label{eq:9}
  \boldsymbol\psi_t(\boldsymbol\omega)=\int \exp\{i\boldsymbol\omega^T\mathbf{y}_{t-1}\}\left(\frac{\partial}{\partial \mathbf{y}_{t-1}}m(\mathbf{y}_{t-1})\right)f(\mathbf{y}_{t-1})d \mathbf{y}_{t-1},
\end{equation}
 where $\boldsymbol\omega\in \mathbb{R}^p$. The quantity $\boldsymbol\psi_t(\boldsymbol\omega)$ is a generalized expectation value of the derivative of the mean function $m(\mathbf{y}_{t-1})$, that is, $\boldsymbol\psi_t(\boldsymbol\omega)$ is the expected value of the gradient of the mean function $m(\mathbf{y}_{t-1})$ weighted by $\exp\{i\boldsymbol\omega^T\mathbf{y}_{t-1}\}$. Moreover, $\boldsymbol\psi_t(\mathbf{0})=\E\left(\frac{\partial}{\partial \mathbf{y}_{t-1}}m(\mathbf{y}_{t-1})\right)$, which is exactly the expected  value of the gradient of $m(\mathbf{y}_{t-1})$. By Theorem 1,  we know that the density-weighted gradient of $m(\mathbf{y}_{t-1})$ spans the TS-CMS, therefore, the Fourier transformation $\boldsymbol\psi_t(\boldsymbol\omega)$ contains all information of the gradient of $m(\mathbf{y}_{t-1})$.  Using the inverse Fourier transformation theorem introduced by Folland (1992), the gradient of $m(\mathbf{y}_{t-1})$ can be computed from $\boldsymbol\psi_t(\boldsymbol\omega)$. Assume $\frac{\partial}{\partial \mathbf{y}_{t-1}}m(\mathbf{y}_{t-1})$ is integrable and continuous on $\mathbb{R}^p$, and $\boldsymbol\psi_t(\boldsymbol\omega)$ is also integrable, then
\begin{align}\label{eq:10}
\left( \frac{\partial}{\partial \mathbf{y}_{t-1}}m(\mathbf{y}_{t-1})\right)f(\mathbf{y}_{t-1})=(2 \pi)^{-p} \int \exp\left\{-i \boldsymbol\omega^T\mathbf{y}_{t-1}\right\} \boldsymbol\psi_t(\boldsymbol\omega) d \boldsymbol\omega.
% \left( \frac{\partial}{\partial \mathbf{x}_{s-1}}m(\mathbf{x}_{s-1} \vert \mathbf{x}_{t-1})\right)f(\mathbf{x}_{s-1}\vert \mathbf{x}_{t-1})=(2 \pi)^{-p} \int \exp\left\{-i \mathbf{x}_{s-1}^T\right\} \boldsymbol\psi_t(\boldsymbol\omega) d \boldsymbol\omega,\label{eq:2.6.2}
\end{align}
The quantities in (\ref{eq:9}) and (\ref{eq:10}) give the relations between $\boldsymbol\psi_t(\boldsymbol\omega)$  and the gradients of $m(\mathbf{y}_{t-1})$, and  based on Theorem \ref{th:1}, the TS-CMS can be spanned by the gradients of $m(\mathbf{y}_{t-1})$. Therefore, $\boldsymbol\psi_t(\boldsymbol\omega)$ can be used to obtain a candidate matrix for the TS-CMS. Some other properties of $\boldsymbol\psi_t(\boldsymbol\omega)$ are given in Proposition \ref{prop:2.1}.

For two time points $t$ and $s$ such that $s>t$, the conditional expectation $y_t$ given $\mathbf{Y}_{t-1}$, and the conditional expectation of $y_s$ given $\mathbf{Y}_{t-1}$ and $\mathbf{Y}_{s-1}$, respectively, are defined as
\begin{align}
%\begin{split}
   m(\mathbf{y}_{t-1}) &= \E [y_t\vert \mathbf{Y}_{t-1}=\mathbf{y}_{t-1}], \label{eq:15} \\
   m(\mathbf{y}_{s-1},\mathbf{y}_{t-1})&= \E[y_s\vert \mathbf{Y}_{s-1}=\mathbf{y}_{s-1},\mathbf{Y}_{t-1}=\mathbf{y}_{t-1}]. \label{eq:16}
%\end{split}
\end{align}
Let $k= s-t$, and if $k < p$, then $\mathbf{Y}_{t-1}$ and $\mathbf{Y}_{s-1}$   have $p-k$  observations in common. 
However, for $k\geq p$, there is no common term, and therefore, we can assume  $\mathbf{Y}_{s-1}$ and $\mathbf{Y}_{t-1} $ are independent of each other.  
 This phenomenon is called $m$-dependence, that is the  time series   $\{y_t\}_{t=1}^N$ is $m$-dependent if $\mathbf{Y}_{s-1}$ and $\mathbf{Y}_{t-1}$ are independent whenever $k-p \geq  m$,   for more details see Lehmann (1999). Therefore, for $s>t$,   we consider two cases as follows

\setlength{\abovedisplayskip}{6.pt}
\setlength{\belowdisplayskip}{6.pt}

\begin{itemize}
     \item \textbf{Case (I):} If $k < p$, then the conditional expectation of $y_s$ given $\mathbf{Y}_{t-1}$ and $\mathbf{Y}_{s-1}$  is  $m(\mathbf{y}_{t-1} , \mathbf{y}_{s-1} $)=$\E[y_s\vert ( \mathbf{Y}_{s-1},\mathbf{Y}_{t-1})]$.
     \item \textbf{Case (II):} If $k \geq p$, then the conditional expectation in   \eqref{eq:16} can be written as $m(\mathbf{y}_{t-1} , \mathbf{y}_{s-1} $)=$m(\mathbf{y}_{s-1} )=\E[y_s\vert \mathbf{Y}_{s-1}]$.
\end{itemize}

\begin{Prop}\label{prop:2.1}
Let $\mathbf{a}(\boldsymbol\omega)$ and $\mathbf{b}(\boldsymbol\omega)$ denote the real and imaginary components of $\boldsymbol\psi_t(\boldsymbol\omega)$.  That is, $\boldsymbol\psi_t(\boldsymbol\omega)=\mathbf{a}(\boldsymbol\omega)+i \mathbf{b}(\boldsymbol\omega)$, then 
\begin{itemize}
	\item[(a)] since $\frac{\partial}{\partial \mathbf{y}_{t-1}}m(\mathbf{y}_{t-1}) \in \mathcal{S}_{\E[y_t\vert \mathbf{Y}_{t-1}]}$, therefore $\mathcal{S}_{\E[y_t\vert \mathbf{Y}_{t-1}]}= \textup{span}\{\mathbf{a}(\boldsymbol\omega),  \mathbf{b}(\boldsymbol\omega$): $\boldsymbol\omega\in  \mathbb{R}^p$\}. That is, $\mathbf{a}(\boldsymbol\omega)$ and $\mathbf{b}(\boldsymbol\omega)$ span the TS-CMS,
	\item[(b)] let $\mathbf{G}(\mathbf{Y}_{t-1})=\frac{\partial}{\partial \mathbf{y}_{t-1}} \log f(\mathbf{Y}_{t-1})$,  and suppose $\log f(\mathbf{y}_{t-1})$ is differentiable such that   $m(\mathbf{y}_{t-1}) f(\mathbf{y}_{t-1})$ goes to zero as $\vert \vert \mathbf{y}_{t-1} \vert \vert \to \infty$, then it can be shown that
\begin{align}\label{eq:12}
  \boldsymbol\psi_t(\boldsymbol\omega)&=-\E_{(\mathbf{Y}_{t-1},y_t)}\left[ y_t\left(\mathbf{G}(\mathbf{Y}_{t-1})+i\boldsymbol\omega\right)\exp\{i\boldsymbol\omega^T\mathbf{Y}_{t-1}\}\right],\\  \label{eq:12-2}
\boldsymbol\psi_s(\boldsymbol\omega)&=-\E_{(\mathbf{Y}_{s-1},y_s)}\left\{[y_s\left(i\boldsymbol\omega+\mathbf{G}(\mathbf{Y}_{s-1}\vert \mathbf{y}_{t-1})\right)\exp\{i\boldsymbol\omega^T\mathbf{Y}_{s-1}\}]\vert \mathbf{y}_{t-1}\right\},
\end{align}

\item[$\mbox{(d)}$]   if   {  $\left(\frac{\partial}{\partial {y}_{t-j}}m(\mathbf{y}_{t-1})\right)f(\mathbf{y}_{t-1})$ and $\left(\frac{\partial}{\partial {y}_{s-j}}m(\mathbf{y}_{s-1}, \mathbf{y}_{t-1})\right)f(\mathbf{y}_{s-1}|\mathbf{y}_{t-1})$} % $\left(\frac{\partial}{\partial \mathbf{y}_{s-j}}m(\mathbf{y}_{s-1})\right)f(\mathbf{y}_{s-1}\vert \mathbf{y}_{t-1},y_t)$
are squared integrable for $1\leq j \leq p$,  and let  $\bar{\boldsymbol\psi}_s(\boldsymbol\omega)$ be  the conjugate of $\boldsymbol\psi_s(\boldsymbol\omega)$,  then, %for all $s$,
\small
	\begin{align}\label{eq:13}
	\begin{split}
	\!\!\!\int \!\! \int \left(\frac{\partial}{\partial\mathbf{y}_{t-1}}m(\mathbf{y}_{t-1})\right)\left(\frac{\partial}{\partial\mathbf{y}_{s-1}}m(\mathbf{y}_{s-1},\mathbf{y}_{t-1})\right)^T &  f(\mathbf{y}_{s-1}\vert \mathbf{y}_{t-1})  f(\mathbf{y}_{t-1}) d\mathbf{y}_{t-1} d\mathbf{y}_{s-1}\\
	&=(2\pi)^{-p} \int \boldsymbol\psi_t (\boldsymbol\omega)\bar{\boldsymbol\psi}_s^T(\boldsymbol\omega) d\boldsymbol\omega.
  \end{split}
      \end{align}
     \normalsize 
\end{itemize}
\end{Prop}

The first assertion of Proposition \ref{prop:2.1} shows that the real and imaginary parts of $\boldsymbol\psi_t(\boldsymbol\omega)$ can be used to generate a candidate matrix for the TS-CMS. The second property represents  $\boldsymbol\psi_t(\boldsymbol\omega)$ as an expectation of a random function that does not include the mean function $m(\mathbf{y}_{t-1})$ explicitly. Therefore, we can estimate $\boldsymbol\psi_t(\boldsymbol\omega)$ without the need of estimating the mean function $m(\mathbf{y}_{t-1})$ and its derivatives directly. This is the main advantage and difference of our approach when compared to the existing methods in the literature that attempt to estimate the mean function $m(\mathbf{y}_{t-1})$ directly, like the NW method (Park et al., 2009). The third property is basically the Riemann-Lebesgue lemma for the Fourier transformation proposed by Folland (1992). It shows that when the density-weighted gradient is absolutely integrable, then $\boldsymbol\psi_t(\boldsymbol\omega)$ converges to zero as the norm of $\boldsymbol\omega$ approaches infinity. The last property of Proposition 3 is a result of applying the Plancherel theorem for the Fourier transform to the density-weighted gradient $\left(\frac{\partial}{\partial \mathbf{y}_{t-1}}m(\mathbf{y}_{t-1})\right)f(\mathbf{y}_{t-1})$ and $\boldsymbol\psi_t(\boldsymbol\omega)$ (Folland 1992, p.  222, 224). In fact, the last property shows the relationship between the expected outer product of  $m(\mathbf{y}_{t-1})$ and  $m(\mathbf{y}_{s-1},\mathbf{y}_{t-1})$  i.e.,  {\small $\E\left[ \left(\frac{\partial}{\partial\mathbf{y}_{t-1}}m(\mathbf{y}_{t-1})\right)\left(\frac{\partial}{\partial\mathbf{y}_{s-1}}m(\mathbf{y}_{s-1},\mathbf{y}_{t-1})\right)^T \right]$},
% {\small$\E\left[\frac{\partial}{\partial \mathbf{Y}_{t-1}}m(\mathbf{Y}_{t-1})\left(\frac{\partial}{\partial \mathbf{Y}_{t-1}}m(\mathbf{Y}_{t-1})\right)^Tf(\mathbf{Y}_{t-1})\right]$},  
and the integral of the outer product of $\boldsymbol\psi_t(\boldsymbol\omega)$ and $\boldsymbol\psi_s(\boldsymbol\omega)$.
\begin{Prop}\label{prop:2.2}
Let $\mathbf{M}_{FMT}^*=(2\pi)^{-p} \int \boldsymbol\psi_t(\boldsymbol\omega)\bar{\boldsymbol\psi}_s^T(\boldsymbol\omega)d\boldsymbol\omega$, then $\mathbf{M}_{FMT}^*$ is a real nonnegative definite matrix, whose column space is exactly equal to the CMS in time series.  That is,  $\mathcal{S}(\mathbf{M}_{FMT}^*)=\mathcal{S}_{\E[y_{t}\vert \mathbf{Y}_{t-1}]}$.
\end{Prop}
We can consider $\mathbf{M}_{FMT}^*$ as the summation of the outer product of $\boldsymbol\psi_t(\boldsymbol\omega)$ over all $\boldsymbol\omega$. From the representation of the density-weighted gradient of $m(\mathbf{y}_{t-1})$ in   \eqref{eq:10}, $\boldsymbol\psi_t(\boldsymbol\omega)$ can be interpreted as the vector of  coefficients of $\exp \{i\boldsymbol\omega^T\mathbf{y}_{t-1}\}$.  Moreover,  $\exp\{i\boldsymbol\omega^T\mathbf{y}_{t-1} \}$ is an oscillatory function of $\boldsymbol\omega$, but not periodic. That is, $\boldsymbol\psi_t(\boldsymbol\omega)$  with smaller $\vert \vert \boldsymbol\omega \vert \vert$ detects lower frequency patterns of $\left(\frac{\partial}{\partial \mathbf{y}_{t-j}}m(\mathbf{y}_{t-1})\right)f(\mathbf{y}_{t-1})$, whereas $\boldsymbol\psi_t(\boldsymbol\omega)$ with larger $\vert \vert \boldsymbol\omega \vert \vert$ captures higher frequency patterns of $\left(\frac{\partial}{\partial \mathbf{y}_{t-j}}m(\mathbf{y}_{t-1})\right)f(\mathbf{y}_{t-1})$. However, based on part(c) of Proposition 3, $\boldsymbol\psi_t(\boldsymbol\omega)$ converges to zero as $\vert \vert \boldsymbol\omega \vert \vert$ goes to infinity. Thus, different values of $\boldsymbol\omega$ produce different patterns with different frequency oscillations in $\boldsymbol\psi_t(\boldsymbol\omega)$. In order to overcome this issue and to obtain a more robust estimator matrix for the TS-CMS, the weight function $W(\boldsymbol\omega)$ is used {to assign difference weights to $\boldsymbol\omega$}. This modification of Proposition \ref{prop:2.2}  is summarized as follows.
\begin{Prop}\label{prop:2.3}
Let $W(\boldsymbol\omega)$ denotes a positive weight function on $\mathbb{R}^p$, and let $Re(\cdot)$ denotes the real part of a complex number, then
\small
\begin{equation}\label{eq:14}
\mathbf{M}_{FMT}=Re\left(\int \boldsymbol\psi_t(\boldsymbol{\omega})\bar{\boldsymbol\psi}_s^T(\boldsymbol\omega) W(\boldsymbol\omega)d\boldsymbol\omega\right),
\end{equation}
\normalsize
 %if $~W(\boldsymbol\omega)$ is a positive function on $\mathbb{R}^p$ then
is a real and nonnegative definite matrix, and $\mathcal{S}(\mathbf{M}_{FMT})=\mathcal{S}_{\E[y_{t}\vert \mathbf{Y}_{t-1}]}$.
\end{Prop}

Proposition \ref{prop:2.3} indicates that the column space of candidate matrix $\mathbf{M}_{FMT}$ can be used to estimate the TS-CMS, i.e., $\mathcal{S}_{\E[y_{t}\vert \mathbf{Y}_{t-1}]}$. The positivity condition of the weight function $W(\boldsymbol\omega)$ in Proposition \ref{prop:2.3} can be relaxed by assuming that $\boldsymbol\psi_t(\boldsymbol\omega)$ is analytic and the weight function has compact support in an open set. However,  in practice, some choices of the weight function will affect the efficiency of the dimension reduction procedure based on $\mathbf{M}_{FMT}$, particularly when the sample size is relatively small. %Similar to Zhu and Zeng (2006),
We employ the Gaussian weight function as $W(\boldsymbol\omega)=(2\pi\sigma_{\omega}^2)^{\frac{-p}{2}}\exp\{-\frac{\vert\vert\boldsymbol\omega\vert\vert^2}{2\sigma_{\omega}^2}\}$, i.e., $W(\boldsymbol\omega)\sim \mathcal{N}(\mathbf{0},\sigma_w^2\mathbf{I}_p)$, where $\sigma_w^2$ is a tuning parameter. Using Propositions \ref{prop:2.1}, \ref{prop:2.2}, and \ref{prop:2.3}, and using the Gaussian weight function $W(\boldsymbol\omega)$, the following explicit expression is obtained for the candidate matrix $\mathbf{M}_{FMT}$. The proof of    \eqref{eq:17} is given in Supplement A.   % \ref{Sup:C}. 
\begin{equation}\label{eq:17}
\mathbf{M}_{FMT}=\E_{(\mathbf{Y}_{t-1},y_t),(\mathbf{Y}_{s-1},y_s)}\Big[\mathbf{J}_{FMT}((\mathbf{Y}_{t-1},y_t),(\mathbf{Y}_{s-1},y_s))\Big],
\end{equation}
\vspace{-.3cm}
where
\vspace{-.3cm}
\small
\begin{align}\label{eq:18}
    & \mathbf{J}_{FMT}((\mathbf{Y}_{t-1},y_t),(\mathbf{Y}_{s-1},y_s))   = \nonumber  \\
    \begin{split}
     &~~   \left\{ \begin{array}{ll}
            y_t y_s \exp\left\{-\frac{\sigma_{\mathbf{\omega}}^2 \vert \vert \mathbf{Y}_{ts} \vert \vert^2}{2} \right\}\left[\sigma^2_{\mathbf{\omega}}\mathbf{I}_p+(\mathbf{G}(\mathbf{Y}_{t-1} )-\sigma^2_{\mathbf{\omega}}\mathbf{Y}_{ts})(\mathbf{G}(\mathbf{Y}_{s-1}\vert \mathbf{Y}_{t-1})+\sigma^2_{\mathbf{\omega}}\mathbf{Y}_{ts})^T\right],~k< p,\\
            y_t y_s \exp\left\{-\frac{\sigma_{\mathbf{\omega}}^2 \vert \vert \mathbf{Y}_{ts} \vert \vert^2}{2} \right\}\left[\sigma^2_{\mathbf{\omega}}\mathbf{I}_p+(\mathbf{G}(\mathbf{Y}_{t-1})-\sigma^2_{\mathbf{\omega}}\mathbf{Y}_{ts})(\mathbf{G}(\mathbf{Y}_{s-1})+\sigma^2_{\mathbf{\omega}}\mathbf{Y}_{ts})^T\right],~~~~~~~~k\geq p,
        \end{array} \right.
     \end{split}
    \end{align}
    \normalsize
where $\mathbf{G}(\mathbf{z})=\frac{\partial}{\partial \mathbf{z}}\log f(\mathbf{z})$, and $\mathbf{Y}_{ts}=\mathbf{Y}_{s-1}-\mathbf{Y}_{t-1}$. 

Suppose $\mathbf{Y}_{t-1}$ follows a multivariate normal distribution, and without loss of generality assume its mean is zero,  i.e.,    $\mathbf{Y}_{t-1} \sim N_p (\mathbf{0},  \boldsymbol\Sigma)$.  Then, it can be shown that     $ \mathbf{G}(\mathbf{y}_{t-1})=\mathbf{y}_{t-1}^T\boldsymbol\Sigma^{-1}$, and $ \mathbf{G}(\mathbf{y}_{s-1})=\mathbf{y}_{s-1}^T\boldsymbol\Sigma^{-1}$, where  $\boldsymbol\Sigma= \left( \gamma_{ij} \right)_{i,j=1}^p$ with $\gamma_{ij}=\gamma(j-i)=\Cov(y_{t-i}, y_{t-j})$.  Moreover, as a result,     $(\mathbf{Y}_{s-1}| \mathbf{Y}_{t-1}=\mathbf{y}_{t-1} ) \sim N_p(\mathbf{0},  \overline{\boldsymbol\Sigma})$ with   $\overline{\boldsymbol\Sigma}={\boldsymbol\Sigma}-\boldsymbol\Sigma_{k}\boldsymbol\Sigma^{-1} \boldsymbol\Sigma_{k} ^T$,  where  $\boldsymbol\Sigma_{k}=\left( \gamma_{ij}(k) \right)_{i,j=1}^p$ with $\gamma_{ij}(k)=\gamma(k+j-i)=\Cov(y_{t-i}, y_{s-j})$.   And it can be shown that  $ \mathbf{G}(\mathbf{y}_{s-1}\vert\mathbf{y}_{t-1})=\mathbf{y}_{s-1}^T\overline{\boldsymbol\Sigma}^{-1}$.  
 It is  assumed $s>t$,  and   $k=s-t$. 
 \iffalse

\textcolor[rgb]{0.00,0.07,1.00}{ Suppose $\mathbf{Y}_{t-1}$ follows a multivariate normal distribution with mean $\mathbf{0}_p$ and variance $\boldsymbol\Sigma$ given by
\begin{equation}\label{eq:sigg}
\boldsymbol\Sigma=
\begin{bmatrix}
\gamma(0)& \dots & \gamma(p-1)\\
\vdots & \vdots &\vdots &\\
\gamma(p-1) & \dots & \gamma(0)
\end{bmatrix},
\end{equation}
where $\gamma(h)=\frac{1}{N-h}\sum_{t=1}^{N-h}y_ty_{t+h}$. Then, $ \mathbf{G}(\mathbf{y}_{t-1})=\mathbf{y}_{t-1}^T\boldsymbol\Sigma^{-1}$, and $ \mathbf{G}(\mathbf{y}_{s-1})=\mathbf{y}_{s-1}^T\boldsymbol\Sigma^{-1}$. Further, suppose $\mathbf{y}_{s-1}\vert \mathbf{y}_{t-1}$ follows a multivariate conditional normal distribution with mean $\mathbf{0}_p$  and conditional variance $\overline{\boldsymbol\Sigma}$ given as
\begin{align}\label{eq:condi}
  \begin{split}
    \overline{\boldsymbol\Sigma}&=\boldsymbol\Sigma-\boldsymbol\Sigma_{s \vert t}\boldsymbol\Sigma^{-1} \boldsymbol\Sigma_{s \vert t} ^T
  \end{split}
\end{align}
where
\begin{equation}\label{eq:sig}
  \boldsymbol\Sigma_{s \vert t}=
\begin{bmatrix}
\gamma(s-t)& \dots & \gamma(s-1-t+p)\\
\vdots & \vdots &\vdots &\\
\gamma(s-p-t+1) & \dots & \gamma(s-t)
\end{bmatrix},
\end{equation}
and $\gamma(h)=\frac{1}{N-h}\sum_{t=1}^{N-h}y_ty_{t+h}$. Then, $ \mathbf{G}(\mathbf{y}_{s-1}\vert\mathbf{y}_{t-1})=\mathbf{y}_{s-1}^T\overline{\boldsymbol\Sigma}^{-1}$.  
 \fi

\subsection{\bfseries Estimation of the Central  Variance Subspace in Time Series}\label{sec:3:2}

Using the same Fourier transform techniques as in the previous section,  we can estimate the time series central variance subspace (TS-CVS).   After removing the conditional mean from the response,  we use a squared residuals-based  Fourier transform method to identify the TS-CVS.   %   From   \eqref{eq:4} 
  Let    $\sigma^2(\mathbf{x}_{t-1})=\E[x_t^2\vert \mathbf{X}_{t-1}=\mathbf{x}_{t-1}]$ represent  the conditional variance  function of $x_t$ in    \eqref{eq:4} given  $\mathbf{X}_{t-1}=(x_{t-1}^2,\cdots,x_{t-q}^2)^T$,  where $x_t=y_t-\E[y_t\vert \mathbf{Y}_{t-1}]$.    From     \eqref{eq:5}, 
  \begin{equation}\label{eq:7.2}
  \sigma^2(\mathbf{x}_{t-1})=\E[x_t^2\vert \mathbf{X}_{t-1}=\mathbf{x}_{t-1}]={g_2(\boldsymbol\Gamma_{d^{'}}^T\mathbf{X}_{t-1})},
\end{equation}
where  $\boldsymbol\Gamma_{d^{'}}=(\Gamma_1,\dots,\Gamma_{d'}) \in \mathbb{R}^{q \times d^{'}}$ ($d^{'}\leq q$),  whose columns form a basis for the TS-CVS.     

Similar to the estimation process of the mean function in Section \ref{sec:3:1},  we use the Fourier transform of $\sigma^2(\mathbf{x}_{t-1})$   to estimate the TS-CVS. 
The practical implementation follows exactly the same steps as those in the previous section,   and therefore for brevity, they are not included here.  
It should be noted that before estimating the TS-CVS,  the impact of the conditional mean of the series needs to be fully removed (Zhu and Zhu,  2009; Park and Samadi,  2020).   Therefore,   we first obtain the residual series $x_t$ by subtracting the estimated mean function from the response ($x_t=y_t-\E[y_t\vert \mathbf{Y}_{t-1}]$); then we use the same Fourier transform method on the squared residuals series to estimate the TS-CVS using candidate matrix $\mathbf{M}_{{FVT}}$.   Unlike Yin and Cook (2002),  and  Ma and Zhu (2013),  this procedure allows us to have a different dimension reduction on the conditional mean and variance functions.

\subsection{\bfseries Estimation of Candidate Matrix }

In this section,  we derive the estimate of  $\mathbf{M}_{FMT}$ %and  $\mathbf{M}_{FVT}$
both under normality and under the more general flexible parametric specification.
 %and discuss its  asymptotic properties.  
 We assume that the dimensionality $d$, the lag order $p$,  and the tuning parameter  $ \sigma_\omega^2$   are known, and discuss their selection in Sections \ref{Sec:5:1}-\ref{Sec:5:3}.

Let $\left\{\mathbf{Y}_{t-1},y_t\right\}_{t=p+1}^{N}$ denote a sequence of random variables and without loss of generality assume $\E[\mathbf{Y}_{t-1}]=\mathbf{0}$. % and $\Cov(\mathbf{Y}_{t-1})=\mathbf{I}_p$.
 Let $F(\mathbf{y}_{t-1},y_t)$ be the cumulative distribution function (cdf) of $(\mathbf{Y}_{t-1},y_t)$. Then, the candidate matrix $\mathbf{M}_{FMT}$ can be represented as
 \small
\begin{equation}\label{eq:21}
  \mathbf{M}_{FMT}=\int \int \mathbf{J}_{FMT}\left((\mathbf{y}_{t-1},y_t),(\mathbf{y}_{s-1},y_s)\right)dF(\mathbf{y}_{t-1},y_t)dF(\mathbf{y}_{s-1},y_s),
\end{equation}
\normalsize
where $\mathbf{J}_{FMT}\left((\mathbf{Y}_{t-1},y_t),(\mathbf{Y}_{s-1},y_s)\right)$ is defined in   (\ref{eq:18}). For the given sequence of random variables $\left\{\mathbf{Y}_{t-1},y_t\right\}_{t=p+1}^{N}$, we use a natural estimator, i.e., the empirical distribution function, to estimate the cdf $F(\mathbf{y}_{t-1},y_t)$ as follows
\begin{equation}\label{eq:22}
  \widehat{{F}}_{N}(\mathbf{y}_{t-1},y_t)=\frac{1}{N-p} \sum_{i=p+1}^{N} \text{I}_{(\mathbf{y}_{i-1}\leq \mathbf{y}_{t-1},y_{i}\leq y_t)},
\end{equation}
where $I_{(\cdot)}$ is the indicator function. Therefore, $\mathbf{M}_{FMT}$ can be estimated by replacing ${F}(\mathbf{y}_{t-1},y_t)$ and ${F}(\mathbf{y}_{s-1},y_s)$ by their corresponding empirical estimators, i.e., $\widehat{{F}}_{N}(\mathbf{y}_{t-1},y_t)$ and $\widehat{{F}}_{N}(\mathbf{y}_{s-1},y_s)$, respectively. Thus, an estimate of the candidate matrix $\mathbf{M}_{FMT}$ is constructed as
\begin{equation}\label{eq:23}
\widehat{\mathbf{M}}_{FMT}=\frac{1}{(N-p)^2}\sum_{t=p+1}^N \sum_{s=p+1}^N \widehat{\mathbf{J}}_{FMT}((\mathbf{y}_{t-1},y_t),(\mathbf{y}_{s-1},y_s)),
\end{equation}
where $\widehat{\mathbf{J}}_{FMT}$ is obtained from ${\mathbf{J}}_{FMT}$ in   (\ref{eq:18}) by replacing $\mathbf{G}(\cdot)$ with $\widehat{\mathbf{G}}(\cdot)$.
% where $\mathbf{G}(\mathbf{z})=\frac{\partial}{\partial \mathbf{z}}\log f(\mathbf{z})$. 
As discussed in  Section \ref{sec:3:1},  under  the assumption of normality,  we have $ \widehat{\mathbf{G}}(\mathbf{y}_{t-1})=\mathbf{y}_{t-1}^T \widehat{\boldsymbol\Sigma}^{-1}$   and  $ \widehat{\mathbf{G}}(\mathbf{y}_{s-1}\vert\mathbf{y}_{t-1})=\mathbf{y}_{s-1}^T\widehat{\overline{\boldsymbol\Sigma}}^{-1}$,  where   
$\widehat{\boldsymbol\Sigma}= \left( \widehat{\gamma}_{ij} \right)_{i,j=1}^p$ with $\widehat{\gamma}_{ij}=\frac{1}{N-h}\sum_{t=h+1}^{N}y_{t-i} y_{t-j}$,  and $h=\max(i,j)$. 
 
In a more general setting, when the density function $f(\mathbf{z})$ is unknown, various nonparametric density estimation methods can be used.  In this work, we use the product of univariate Gaussian kernel functions to estimate the multivariate density function $f(\mathbf{z})$. The nonparametric kernel function is defined similar to that  in Park et al. (2009) as follows
\small
\begin{equation}\label{eq:19}
\widehat{f}(\mathbf{z})=\left(N \prod_{i=1}^r a_{Ni}\right)^{-1}\sum_{j=1}^{N} \prod_{i=1}^r K\left(\frac{z_i-X_{ij}}{a_{Ni}}\right),
\end{equation}
\normalsize
where $\mathbf{z}=\left(z_1,z_2,\dots,z_r \right)^T$,  $\mathbf{X}=\left(X_1,X_2,\dots,X_r \right)^T$,  and $K(\cdot)$ is a univariate Gaussian kernel function with bandwidth $a_{Ni}=b_rs_iN^{-\frac{1}{4+r}}$, for $i=1,\dots,r$, with $b_r=\left(\frac{4}{r+2}\right)^{\frac{1}{r+4}}$, and $s_i$ is the corresponding sample standard deviation of $X_i$. The constant $b_r$ in $a_{Ni}$ is chosen as suggested in Silverman (1986) and Scott (1992).\\
When $k= s-t<p$,  the quantity $\mathbf{G}(\mathbf{Y}_{s-1}\vert \mathbf{Y}_{t-1})$ in   (\ref{eq:18}) can be written as follows 
%Let $k=\vert t-s \vert $, then,
\small
\begin{align}\label{eq:20}
%\begin{split}
  \mathbf{G}(\mathbf{Y}_{s-1}\vert \mathbf{Y}_{t-1}) &= \frac{\partial}{\partial \mathbf{Y}_{s-1}}  \log \left(f(\mathbf{Y}_{s-1}\vert \mathbf{Y}_{t-1}) \right)
  %\textcolor[rgb]{0.00,0.07,1.00}{=\frac{\frac{\partial}{\partial \mathbf{y}_{t-1}}f(\mathbf{Y}_{s-1}\vert \mathbf{Y}_{t-1})}{f(\mathbf{Y}_{s-1}\vert \mathbf{Y}_{t-1})}}  \\
  = \frac{\partial}{\partial \mathbf{Y}_{s-1}}  \log \left( \frac{f(\mathbf{Y}_{s-1},\mathbf{Y}_{t-1})}{f(\mathbf{Y}_{t-1})}	\right) \\
%&= \frac{\partial}{\partial \mathbf{Y}_{s-1}}  \log \left( f(\mathbf{Y}_{s-1},\mathbf{Y}_{t-1}) \right)-  \frac{\partial}{\partial \mathbf{Y}_{s-1}} \log\left(f(\mathbf{Y}_{t-1}) \right)\\  
=&\frac{\partial}{\partial \mathbf{Y}_{s-1}}  \log \left( f(\mathbf{Y}_{s-1},\mathbf{Y}_{t-1}) \right)- \left[\mathbf{0}_{1\times k} \hspace{0.05cm} \left(\frac{\partial}{\partial \text{y}_{s-k-1}} \dots \frac{\partial}{\partial \text{y}_{s-p}} \right) \log\left(f(\mathbf{Y}_{t-1}) \right)\right]^T.  \nonumber
%\end{split}
\end{align}
\normalsize
Here,  we need to estimate both the gradient of the log-joint density of $\mathbf{Y}_{t-1}$ and $\mathbf{Y}_{s-1}$,  and the gradient of the log-density of $\mathbf{Y}_{t-1}$ with respect to  $\mathbf{Y}_{s-1}$.
We have the following justifications for the kernel function in  (\ref{eq:19}) based on two scenarios.  For  $s>t$ and $k=s-t$,  when $k< p$ (Case (I)),  we set $r=p+k$   in   (\ref{eq:19}) to estimate $f(\mathbf{y}_{t-1},\mathbf{y}_{s-1})$.   However, when $k\geq p$ (Case (II)),  then we have $\mathbf{G}(\mathbf{Y}_{s-1}\vert \mathbf{Y}_{t-1})=\mathbf{G}(\mathbf{Y}_{s-1})$, therefore we set $r=p$ in   (\ref{eq:19}).

In particular,  let $\widehat{\mathbf{G}}(\mathbf{Y}_{t-1})=\left( \widehat{G}_{1} (\mathbf{Y}_{t-1}), \dots, \widehat{G}_{p} (\mathbf{Y}_{t-1}) \right)^T$,  and  $a_{N_i}=a_N, i=1,\dots, p$. Then,  by   applying the  kernel density estimation given in   \eqref{eq:19},   it can be shown that   
\small
\begin{align}\label{eq:5.00}
\begin{split}
\widehat{G}_{i} (\mathbf{Y}_{t-1})&=\frac{\nabla_{{y}_{t-i}}\widehat{f}(\mathbf{Y}_{t-1})}{\widehat{f}(\mathbf{Y}_{t-1})} =\frac{-y_{t-i}}{a_N^2}+ \frac{\sum_{r=p+1}^{N} \frac{y_{r-i}}{a_N^2} \exp\left\{ -\sum_{j=1}^{p} \frac{\left(y_{t-j}-y_{r-j} \right)^2}{2a^2_{N}} \right\}}{\sum_{r=p+1}^{N}\exp\left\{-\sum_{j=1}^{p} \frac{\left( y_{t-j}-y_{r-j} \right)^2 }{2a^2_{N}} \right\}}.
\end{split}
\end{align}
\normalsize

Similarly,    from    \eqref{eq:20} for  $\widehat{\mathbf{G}}(\mathbf{Y}_{s-1}|\mathbf{Y}_{t-1})=\left( \widehat{G}_{1} (\mathbf{Y}_{s-1} | \mathbf{Y}_{t-1}), \dots, \widehat{G}_{p} (\mathbf{Y}_{s-1} | \mathbf{Y}_{t-1}) \right)^T$,  with $s>t$ and  $k=s-t$,     it can be shown that 
 \begin{equation}\label{eq:5.001}
 \widehat{G}_{i} (\mathbf{Y}_{s-1} | \mathbf{Y}_{t-1})=  \widehat{G}_{i}^{A}(\mathbf{Y}_{s-1} | \mathbf{Y}_{t-1})- \widehat{G}_{i}^{B}(\mathbf{Y}_{s-1} | \mathbf{Y}_{t-1}), ~~~~ i=1, \dots, p, 
 \end{equation}
 \vspace{-.3cm}
where 
\small
\begin{align*}%\label{eq:11g}
   \widehat{G}_{i}^{A}(\mathbf{Y}_{s-1} | \mathbf{Y}_{t-1})=  \frac{\nabla_{{y}_{s-i}}\widehat{f}(\mathbf{Y}_{s-1}, \mathbf{Y}_{t-1})}{\widehat{f}(\mathbf{Y}_{s-1}, \mathbf{Y}_{t-1})} &=\frac{-{y}_{s-i}}{a_N^2}+\frac{ \sum_{r=p+1}^{N}\left(\frac{{y}_{r-i}}{a_N^2}\right)\exp\left\{-\sum_{j=1}^{p+k}\frac{\left(y_{s-j}-y_{r-j}\right)^2}{2a_N^2}\right\}
     }{\sum_{r=p+1}^{N} \exp\left\{-\sum_{j=1}^{p+k}\frac{\left(y_{s-j}-y_{r-j}\right)^2}{2a_N^2}\right\}},
\end{align*}
\normalsize
and $\widehat{G}_{i}^{B}(\mathbf{Y}_{s-1} | \mathbf{Y}_{t-1})=0$,  if $i\leq k$,  otherwise,  i.e.,  when $i>k$, it is given by 
{\small
\begin{align*}%\label{eq:14g}
  \begin{split}
 \widehat{G}_{i}^{B}(\mathbf{Y}_{s-1} | \mathbf{Y}_{t-1})=    \frac{\nabla_{{y}_{s-i}}\widehat{f}(\mathbf{Y}_{t-1})}{\widehat{f}(\mathbf{Y}_{t-1})} 
     &=\frac{-y_{s-i} }{a_N^2}+\frac{\sum_{r=p+1}^{N}\frac{y_{r-i}}{a_N^2}\exp\left\{-\sum_{j=1}^{p}\frac{\left(y_{t-j}-y_{r-j}\right)^2}{2a_N^2}\right\}}{\sum_{r=p+1}^{N} \exp\left\{-\sum_{j=1}^{p}\frac{\left(y_{t-j}-y_{r-j}\right)^2}{2a_N^2}\right\}}.
  \end{split}
\end{align*}
\normalsize}

 The density function of $\mathbf{y}_{t-1}$ is appeared in the denominator of $\mathbf{G}(\mathbf{y}_{t-1})$,  and the conditional density function of $\mathbf{y}_{s-1}$ given $\mathbf{y}_{t-1}$ is appeared in the denominator of $\mathbf{G}(\mathbf{y}_{s-1}\vert \mathbf{y}_{t-1})$, {see   \eqref{eq:20}}.  To avoid the negative impact  of small values of $\widehat{f}(\mathbf{y}_{t-1})$ and $\widehat{f}(\mathbf{y}_{s-1}\vert \mathbf{y}_{s-1})$, we exclude  them by imposing a predetermined threshold as $ \widehat{I}_{t}=I_{\hat{f}(\mathbf{y}_{t-1})>c_N}$ and  $ \widehat{I}_{s}=I_{\hat{f}(\mathbf{y}_{s-1}\vert \mathbf{y}_{t-1})>c_N}$,   
where $c_N$ is a predefined threshold. Then, the updated estimator of $\mathbf{M}_{FMT}$ becomes
\begin{equation}\label{eq:25}
  \widehat{\mathbf{M}}_{FMT}=\frac{1}{(N-p)^2} \sum_{t=p+1}^{N}\sum_{s=p+1}^{N} \widehat{\mathbf{J}}_{FMT}(\mathbf{z}_t,\mathbf{z}_s)\hat{I}_{t}\hat{I}_{s},
\end{equation}
where $\mathbf{z}_t=(\mathbf{y}_{t-1},y_t)$ and $\mathbf{z}_s=(\mathbf{y}_{s-1},y_s)$.  To guarantee the asymptotic consistency of the proposed estimator $\widehat{\mathbf{M}}_{FMT}$ in   (\ref{eq:25}), it is required $c_N \longrightarrow 0$ as $N \longrightarrow \infty$. In the subsequent sections, we use $c$ to represent $c_N$.
%%%%%%%%%%%%%%%%%%%%%%%%%%%%%%%%%%%%%%%%%%%%%%%%%%%%%%% SECTION 3 %%%%%%%%%%%%%%%%%%%%%%%%%%%%%%%%%%%%%%%%%%%%%%%%%%%%%%%%%%%%%%%%%%%%%%%%%%%%%

\section{ \large  Consistency of the Estimators}

In this section,  the asymptotic properties of the proposed estimators are established under two scenarios: one under normality and the other under general assumptions. 
The asymptotic results are obtained by using the dependent U-statistic and dependent von-Mises-Statistic (V-Statistic), which are defined in Definition \ref{def:4.1}.
\begin{Definition}\label{def:4.1}
  Let $\mathbf{y}_{t-1} \in \mathbb{R}^{p}$, and let $\mathbf{h}: \left( \mathbb{R}^p \right)^m \to \mathbb{R}^{p\times p}$  is  a measurable symmetric
function of its arguments $\mathbf{y}_{t-1} \in \mathbb{R}^{p}$. Then, the U-statistic $\mathbf{U}_N(\mathbf{h})$ and the von-Mises-statistic (V-statistic),   $\mathbf{V}_N(\mathbf{h})$, respectively, are defined as
\small
\begin{align}
 \mathbf{U}_N(\mathbf{h})&=\binom{N}{m}^{-1}\sum_{1\leq t_1< \dots < t_m\leq N} \mathbf{h}(\mathbf{Y}_{t_1},\dots,\mathbf{Y}_{t_m}), \label{eq:57}\\
 \mathbf{V}_N(\mathbf{h})&=N^{-m}\sum_{t_1=1}^{N}\dots\sum_{t_m=1}^{N} \mathbf{h}(\mathbf{Y}_{t_1},\dots,\mathbf{Y}_{t_m}),\label{eq:58}
\end{align}
\normalsize
where  $\mathbf{h}(\cdot)$ is 
%a symmetric function, i.e., $h(\dots,\mathbf{y}_{t_i},\dots,\mathbf{y}_{t_j},\dots)=h(\dots,\mathbf{y}_{t_j},\dots,\mathbf{y}_{t_i},\dots)$,  for all $\mathbf{y}_{t_i}, \mathbf{y}_{t_j}\in \mathbb{R}^{p}$, and 
called the kernel of the U- and V-statistics (see Dehling, 2006).  
\end{Definition}
 The following lemma summarizes the asymptotic property of dependent U- and V- statistics (Dehling, 2006).  The proof of Lemma \ref{lem:1} is given in Supplement A.  % \ref{Sup:C}. 
    \begin{lem}\label{lem:1}
      Suppose $\{\mathbf{Y}_t\}_{t=p+1}^{N}$ be a stationary ergodic process of $\mathbb{R}^p$-valued random variables. %Let $h:\left(\mathbb{R}^{p}\right)^m \to \mathbb{R}$ be a measurable symmetric function, % where $h(\cdot)$ is symmetric under the permutation of its $m$ variables,
     % i.e., $h(\dots,\mathbf{y}_{t_i},\dots,\mathbf{y}_{t_j},\dots)=h(\dots,\mathbf{y}_{t_j},\dots,\mathbf{y}_{t_i},\dots)$.
      %That is, $\mathbf{P}(\mathbf{Y}_i,\mathbf{Y}_j)=\mathbf{P}(\mathbf{Y}_j,\mathbf{Y}_i)$ for all $i \neq j$ and $1\geq i,j\leq m$.
      %Then, we can defined the U-statistics and V-statistics as follows,
      %\begin{equation}\label{eq:3.1}
      %U_N(\mathbf{P})=\binom{N}{m}^{-1} \sum_{1<t_1<\dots<t_m}\mathbf{P}(\mathbf{Y}_1,\dots,\mathbf{Y}_{m}),
      %\end{equation}
      %\begin{equation}
      %V_N(\mathbf{P})=\frac{1}{N^2}\sum_{t_1<t_2}\dots \sum_{t_{m-1}<t_m} \mathbf{P}(\mathbf{Y}_1,\dots,\mathbf{Y}_{m}).
      %\end{equation}
      Consider a general U- and V-statistics as in Definition \ref{def:4.1}.
      %\begin{align}\label{eq:27}
      %\begin{split}
       %           {U}_N(h) &= \binom{N}{m}^{-1}\sum_{1\leq t_1< \dots < t_m \leq N} h(\mathbf{Y}_{t_1},\dots,\mathbf{Y}_{t_m}),\\
       %           {V}_N(h) &= N^{-m}\sum_{t_1=1}^{N} \dots \sum_{t_m=1}^{N}h(\mathbf{Y}_{t_1},\dots,\mathbf{Y}_{t_m}),
      %\end{split}
      %\end{align}
        %Let
       %$\hat{\mathbf{u}}_N=\theta_N+\frac{m}{N}\sum_{i=1}^{N}[\mathbf{r}_N(\mathbf{y}_i)-\boldsymbol\theta_N]$, where
       %$\boldsymbol\theta_N=\E[\mathbf{P}_N(\mathbf{Y}_1,\dots,\mathbf{Y}_m)]$.
       %and $\mathbf{r}_N(\mathbf{y}_i)=\E[\mathbf{P}_N(\mathbf{Y}_1,\dots,\mathbf{Y}_m)\vert \mathbf{Y}_i=y_i]$.
       If $\E[\vert \vert \mathbf{h}(\mathbf{Y}_{t_1},\dots,\mathbf{Y}_{t_m})\vert \vert ^2]=o_p(N)$ for any $t_{1},\dots,t_m \in \{1,\dots,m\}$, then $\sqrt{N}(\mathbf{V}_N(\mathbf{h})-\widehat{\mathbf{U}}_N(\mathbf{h}))=o_p(1)$ where $\widehat{\mathbf{U}}_N(\mathbf{h})=\boldsymbol\theta+\frac{m}{N}\sum_{i=p+1}^{N} \mathbf{h}_1(\mathbf{Y}_{t_i})$, $\boldsymbol\theta=\E[\mathbf{h}(\mathbf{Y}_{t_1},\dots,\mathbf{Y}_{t_m})]$,  and $\mathbf{h}_1(\mathbf{Y}_{t_i})=\E[\mathbf{h}(\mathbf{Y}_{t_1},\dots,\mathbf{Y}_{t_m})\vert \mathbf{Y}_{t_i}]-\boldsymbol\theta$.
       %\begin{itemize}
        %     \item [(i),] $\frac{\sqrt{N}(\mathbf{U}_N-\theta_N)}{4\sigma^2}=o_p(1)$.
         %    \item [(ii),] $\sqrt{N}(\mathbf{V}_N-\mathbf{U}_N)=o_p(1)$.
       %\end{itemize}
    \end{lem}

\subsection{\bfseries Asymptotic Properties Under Normality }
    In this part,  the asymptotic behaviors of the estimators are obtained under the normal assumption.  
%As it is discussed in  Section \ref{sec:3:1},    when   $\mathbf{Y}_{t-1} \sim N_p (\mathbf{0},  \boldsymbol\Sigma)$,  then   $(\mathbf{Y}_{s-1}| \mathbf{Y}_{t-1}=\mathbf{y}_{t-1}) \sim N_p(\mathbf{0},  \overline{\boldsymbol\Sigma})$,  where  $\overline{\boldsymbol\Sigma}=\boldsymbol\Sigma-\boldsymbol\Sigma_{k}\boldsymbol\Sigma^{-1} \boldsymbol\Sigma_{k} ^T$. 
%And, it can be shown that    $ \mathbf{G}(\mathbf{y}_{t-1})=\mathbf{y}_{t-1}^T\boldsymbol\Sigma^{-1}$,   $ \mathbf{G}(\mathbf{y}_{s-1})=\mathbf{y}_{s-1}^T\boldsymbol\Sigma^{-1}$,  and   $ \mathbf{G}(\mathbf{y}_{s-1}\vert\mathbf{y}_{t-1})=\mathbf{y}_{s-1}^T\overline{\boldsymbol\Sigma}^{-1}$. 
   We use $\mathbf{J}_{FMTn}$, $\mathbf{M}_{FMTn}$, and $\widehat{\mathbf{M}}_{FMTn}$ to respectively  denote $\mathbf{J}_{FMT}$, $\mathbf{M}_{FMT}$, and $\widehat{\mathbf{M}}_{FMT}$ under the normality assumption.
\begin{theorem}\label{thm:normal}
Suppose $\mathbf{Y}_{t-1}$ follows multivariate normal distribution,  and let    $\mathbf{Z}_t=(\mathbf{Y}_{t-1},y_t)$ and $\mathbf{Z}_s=(\mathbf{Y}_{s-1},y_s)$,  and assume   the  covariance matrix of $\vect\left(\mathbf{J}_{FMTn}(\mathbf{Z}_{t},\mathbf{Z}_{s})\right)$ exists. As $N \to \infty$,  we have
  \begin{equation*}%\label{eq:Normal}
    \widehat{\mathbf{M}}_{FMTn}=\mathbf{M}_{FMTn}+\frac{1}{N}\sum_{t=p+1}^{N}\left(\mathbf{J}_{FMTn}^{(1)}
    (\mathbf{z}_{t})-2\mathbf{M}_{FMTn}\right)+o_p(N^{-1/2}),
  \end{equation*}
where $ \mathbf{J}_{FMTn}^{(1)} (\mathbf{z})=\E_{\mathbf{Z}_{s}}\left[\mathbf{J}_{FMTn}\left(\mathbf{z},\mathbf{Z}_{s}\right)+ \mathbf{J}^T_{FMTn}\left(\mathbf{z},\mathbf{Z}_{s}\right) \right]$. 
%\begin{equation*}%\label{eq:norml2}
%  \mathbf{J}_{FMTn}^{(1)} (\mathbf{z})=\E_{\mathbf{Z}_{s}}\left[\mathbf{J}_{FMTn}\left(\mathbf{z},\mathbf{Z}_{s}\right)+ \mathbf{J}^T_{FMTn}\left(\mathbf{z},\mathbf{Z}_{s}\right) \right].
%\end{equation*}
Let $\boldsymbol\Sigma_{FMTn}$ be the covariance matrix of $\vect\left(\mathbf{J}_{FMTn}^{(1)}(\mathbf{Z}_{t})\right)$. Then,  as  $N \to \infty$, 
\begin{equation*}%\label{eq:norml3}
  \sqrt{N}\left(\vect(\widehat{\mathbf{M}}_{FMTn}) -\vect(\mathbf{M}_{FMTn})\right) \overset{\mathcal{D}}\longrightarrow \mathcal{N}(\mathbf{0},\boldsymbol\Sigma_{FMTn}).
\end{equation*}
\end{theorem}

%\textcolor[rgb]{0.00,0.07,1.00}{.
%\textcolor[rgb]{0.00,0.07,1.00}{ 
\subsection{\bfseries Asymptotic Properties Under General Assumptions}
In this section, the asymptotic properties of the proposed estimator of the TS-CMS are derived under some general assumptions.  These assumptions are required when $\mathbf{Y}_{t-1}$  does not follow a normal distribution.  The proof of Theorem \ref{th:2.2} is given in Supplement A. % \ref{Sup:C}.

\begin{theorem}\label{th:2.2}
  Let $a_{N}$ be the bandwidth of the kernel function in   (\ref{eq:19}). Suppose assumptions (A1)-(A9) in the  Supplement A %\ref{Sup:C} 
   hold.  If,
  \begin{itemize}
    \item [(a)] $a_{N} \to 0$ as $N \to \infty$,and $\left(b^{-1}a_{N}/\log N \right) \to 0$, as $b \to 0$;
    \item [(b)] for some $\epsilon>0$, $\left(b^4N^{1-\epsilon}a_{N}^{2p+2} /\log N\right) \to \infty$; and ~~~~~~~~~ (c)  $Na_{N}^{2\ell-2} \to 0$,
  %  \item [(c)] and $Na_{N}^{2\ell-2} \to 0$,
  \end{itemize}
  then,  $\vect(\widehat{\mathbf{M}}_{FMT})$ asymptotically follows a multivariate normal distribution as
  \begin{equation}\label{eq:28}
    \sqrt{N}\left(\vect(\widehat{\mathbf{M}}_{FMT})-\vect(\mathbf{M}_{FMT})\right)\xrightarrow[]{{\mathcal{D}}}  \mathcal{N} \left(\mathbf{0},\boldsymbol\Sigma_{FMT}\right),
  \end{equation}
  where, $\boldsymbol{\Sigma}_{FMT}$ is the covariance matrix of $\vect\left( \mathcal{R}(\mathbf{z}_s) + \mathcal{R}^T(\mathbf{z}_s) \right)$,  where 
  \small 
 \begin{equation*} 
    \mathcal{R}(\mathbf{z}_s)\!\!=\! \! \! \int \!\! \E_{\mathbf{Z}_t} \!\! \left[ y_t \boldsymbol\phi(\mathbf{Y}_{t-1},   \boldsymbol\omega) \left([y_{s}-m(\mathbf{y}_{s-1},\mathbf{Y}_{t-1})]\boldsymbol\phi(\mathbf{y}_{s-1}|\mathbf{Y}_{t-1},   \boldsymbol\omega)- \! \mathbf{U}(\mathbf{Y}_{t-1}, \mathbf{y}_{s-1},\boldsymbol\omega ) \right)^{T}  \! W(\boldsymbol\omega)\right]\!d\boldsymbol\omega,
   \end{equation*}
   \normalsize 
% \begin{equation*}
%     \mathcal{R}(\mathbf{z}_r) =\int \boldsymbol\psi_s(\boldsymbol\omega)\left\{[m(\mathbf{y}_{t-1})-y_{t}]\phi(\mathbf{y}_{t-1},\boldsymbol\omega)+\frac{\partial m(\mathbf{y}_{t-1})}{\partial \mathbf{y}_{t-1}}H(\mathbf{y}_{t-1},\boldsymbol\omega)\right\}^{T}d\boldsymbol\omega,
%   \end{equation*}  
%  and
%$\phi(\mathbf{y}_{t-1},\boldsymbol\omega)=\frac{\partial}{\partial \mathbf{y}_{t-1}}H(\mathbf{y}_{t-1},\boldsymbol\omega)+H(\mathbf{y}_{t-1},\boldsymbol\omega)\frac{\partial}{\partial \mathbf{y}_{t-1}} \log f(\mathbf{y}_{t-1})$.
with 
 $\mathbf{U}(\mathbf{y}_{t-1}, \mathbf{y}_{s-1},\boldsymbol\omega )= \left( \frac{\partial  }{\partial \mathbf{y}_{s-1}} m(\mathbf{y}_{s-1},\mathbf{y}_{t-1}) \right) H(\mathbf{y}_{s-1},\boldsymbol\omega)$,   $\boldsymbol\phi(\mathbf{y}_{t-1},   \boldsymbol\omega)=\frac{\partial}{\partial \mathbf{y}_{t-1}}H(\mathbf{y}_{t-1},\boldsymbol\omega)\\+H(\mathbf{y}_{t-1},\boldsymbol\omega)\mathbf{G}( \mathbf{y}_{t-1})$,  and 
$\boldsymbol\phi(\mathbf{y}_{s-1}|\mathbf{y}_{t-1},  \boldsymbol\omega)=\frac{\partial}{\partial \mathbf{y}_{s-1}}H(\mathbf{y}_{s-1},\boldsymbol\omega) +H(\mathbf{y}_{s-1},\boldsymbol\omega)\mathbf{G}(\mathbf{y}_{s-1}\vert \mathbf{y}_{t-1})$.
\end{theorem}

%%%%%%%%%%%%%%%%%%%%%%%%%%%%%%%%%%%%%%%%%%%%%%%%%%%%%%%%%%%%%%%%%%%%%%%%%%%%%%%%%%%%%%%%%%%%%%%%%%%%%%%%%%%%%%
\section{\large  Algorithms to Estimate the Time Series   CMS}\label{Sec:5}
In this section, we describe the process of estimating the time series central mean subspace (\text{TS-CMS}). An estimate of the candidate matrix ${\mathbf{M}}_{FMT}$, whose column space recovers the TS-CMS, is proposed in  \eqref{eq:23}. We first discuss the estimation procedures that are used to estimate the unknown lag order $p$, the dimension of the TS-CMS $d$, and the tuning parameter $\sigma_w^2$. Then, we describe the process of estimating the \text{TS-CMS}.

To estimate the unknown lag order $p$, and the dimension $d$ of the \text{TS-CMS},  we apply the bootstrapping procedure proposed by Ye and Weiss (2003). Notice that the classical bootstrap methods cannot be used for time series data since they are designed for independent and identically distributed (iid) observations. Therefore, the traditional bootstrapping approach does not preserve the ordering and dependencies between the time series observations. To overcome this problem, we instead use the {\it block bootstrapping} approach. The main idea of the block bootstrapping method is to sample the whole block instead of a single observation and then concatenate them together in such a way that preserves the original time series structure. There are several methods for doing block bootstrapping, such as fixed block bootstrapping, stationary block bootstrapping, moving block bootstrapping, and model-based block bootstrapping techniques (H\"{a}rdle et al., 2003).
%We employ a fixed-block bootstrapping procedure with a block size of $N/2$, where $N$ is the sample size.%Here, we use a fixed block bootstrapping sampling procedure with a fixed block size of $N/2$, where $N$ is the sample size.  %For more details on time series bootstrapping, see H\"{a}rdle  et al. (2003).

\subsection{\bfseries Algorithm for estimating p and d}\label{Sec:5:1}
Let $\{y_t\}_{t=1}^N$ be a time series process and set $\mathbf{Y}_{t-1}=(y_{t-1},\dots,y_{t-p})^T$. Recall that $p$ is the number of lags, and $d$ is the number of directions in the time series model defined in   (\ref{eq:4}) (dimension of the TS-CMS), where $d \leq p$. Moreover, let $\mathcal{\widehat{S}}(p,d)$ denotes the estimation of the TS-CMS $\mathcal{S}_{\E[y_t\vert \mathbf{Y}_{t-1}]}$, where $p$ and $d$ are known. Then, the variability of $\mathcal{\widehat{S}}(p,d)$  can be calculated by the following four steps fixed block bootstrap procedure:
\begin{itemize}
  \item[1.] Perform a fixed block resampling of the time series observation $\{y_t\}_{t=1}^N$ to generate $B$ bootstrap  samples of size $N$, and denote the $b$th  sample by $\{y_t^{(b)}\}_{t=1}^N, b=1,\dots,B$.
  \item[2.] For each of the bootstrap time series samples $\{\mathbf{Y}^{(b)}_{t-1}, y_t^{(b)}\}_{t=p+1}^N$ and given $p$ and $d$, estimate $\mathcal{S}_{\E[y_t\vert \mathbf{Y}_{t-1}]}$, and denote it by $\mathcal{\widehat{S}}^{(b)}(p,d)$.
  \item[3.] Then, calculate the distance between $\mathcal{\widehat{S}}^{(b)}(p,d)$ and $\mathcal{\widehat{S}}(p,d)$, and show it by $D^{(b)}(p,d)$.
  \item[4.]  Calculate the mean distance between {\small$\mathcal{\widehat{S}}^{(b)}(p,d)$ and $\mathcal{\widehat{S}}(p,d)$} over all $B$  bootstrap samples  
\begin{equation}\label{eq:29}
  \bar{D}(p,d)=\frac{1}{B}\sum_{j=1}^BD^{(b)}(p,d).
\end{equation}
\end{itemize}
In practice, however, $p$ and $d$ are unknown and need to be estimated. To this end, we repeat the above four steps for all combinations of a predetermined set of possible candidates of $p$, i.e., $\{p_1,\dots,p_{\kappa}\}$, and $d_i~(\leq p_i$ for $i=1,\dots,\kappa$). Therefore, for each $p_i, i=1,\dots,\kappa$, we have a sequence of estimators $\{\bar{D}(p_i,d_i)\}_{d_i=1}^{p_i}$.
In order to achieve dimension reduction for a given $p_i$, the estimator $\hat{d}_i$ is chosen to minimize the variability measure $\{\bar{D}(p_i,d_i)\}_{d_i=1}^{p_i}$ and satisfy the condition $\hat{d}_i < p_i$. For each lag order $p_i$, we obtain the corresponding estimator $\hat{d}_i$. Then, the estimation of $(\hat{p},\hat{d})$ is the pair that gives the minimum variability among all the possible pairs of $(p_i,\hat{d}_i),~\hat{d}_i< p_i$, $i=1,\dots,\kappa$.
%The estimator $\hat{d}$ for each candidate $p$ is obtained as the valley point explain in Zhu and Zeng(2006). That is, suppose true dimension of $\mathcal{S}_{\E[y_t\vert \mathbf{Y}_{t-1}]}$ is $d_0$.  Then $\bar{d}(p,q)$  is shows the following trend. When $1\leq d \leq d_0$, $\bar{d}(p,q)$ is decreases and then increases from $d_0\leq d \leq d_*$, where $d_*$ is the maximizer of $\bar{d}(p,q)$, and then decreases to zero when $d_*\leq d \leq p$. Then $d_0$ known as the valley point. In other word, $\hat{d}$ will be the first minimum between $1\leq d\leq d^*$ where $d^*$ is the maximizer of the $\bar{d}(p,q)$.

\subsection{\bfseries Algorithm to estimate the tuning parameter $\sigma_w^2$} \label{Sec:5:2}
\setlength{\abovedisplayskip}{6.pt}
\setlength{\belowdisplayskip}{6.pt}

The tuning parameter $\sigma_w^2$ can be considered as the bandwidth of the weight function $W(\boldsymbol\omega)$ introduced in Proposition \ref{prop:2.3}, as $W(\boldsymbol\omega)=(2\pi \sigma_w^2)^{-1/2}\exp\{-0.5\vert\vert \boldsymbol\omega\vert\vert^2/\sigma_w^2\}$. However, it is different from the bandwidth parameter in kernel density estimation. Unlike in kernel density estimation, in which the consistency of the estimator requires the bandwidth to converge to zero as the sample size goes to infinity, here the estimator $\widehat{\mathbf{M}}_{FMT}$ is consistent for any positive value of $\sigma_w^2$ (see Proposition \ref{prop:2.3}). However, both small and large values of $\sigma_w^2$ are problematic. Therefore, the choice of $\sigma_w^2$ affects the estimates of the TS-CMS, thus it should be chosen carefully.  When $\sigma_w^2$ is very small and around zero, the weight function assigns small weights to $\boldsymbol\psi_t(\boldsymbol\omega)$ in   (\ref{eq:14}), and as a result, some of the symmetric directions may be missed. On the other hand, when $\sigma_w^2$ is too large, then larger values of $\vert \vert \boldsymbol\omega \vert \vert$ would assign larger weights to $\boldsymbol\psi_t(\boldsymbol\omega)$. The larger weights of $\boldsymbol\psi_t(\boldsymbol\omega)$ correspond to the higher frequency patterns and cause to increase in the noises in the estimation, this makes the estimator  $\widehat{\mathbf{M}}_{FMT}$ biased unstable. %Our empirical studies show that, $\sigma_w^2=0.1$ generally performs well to estimate the pair $(p,d)$, \textcolor[rgb]{1.00,0.00,0.00}{however it does not affect much on the estimation of the TS-CMS because a positive $\sigma_w^2$ is sufficient to obtain the candidate matrix $\mathbf{M}_{FMT}$.} Notice that, $\sigma_w^2=0.1$ is recommended by Zhu and Zeng (2006) in regression model.
For improving the accuracy of estimation in the time series model, it is better to estimate $\sigma_w^2$ for each time series before estimating the TS-CMS. The following bootstrap scheme is used to choose the optimal value of $\sigma_w^2$.

Let $(\sigma_{w,1}^2,\dots,\sigma_{w,\kappa}^2)$ represents a grid of candidate values  for $\sigma_w^2$ that are equally spaced on a given interval. Moreover, for a given time series $\{y_t\}_{t=1}^N$, and given $\sigma_{w,i}^2$, $p$, and $d$, let $\mathcal{\widehat{S}}({\sigma_{w,i}^{2}})$ denotes the corresponding estimate of $\mathcal{S}_{\E[y_t\vert \mathbf{Y}_{t-1}]}$. Then, an appropriate value of $\sigma_w^2$ can be obtained by using the block bootstrap procedure as follows:
\begin{itemize}
  \item[1.] Perform a fixed block bootstrap resampling of the time series $\{y_t\}_{t=1}^N$ to generate $B$ bootstrap  samples of size $N$, and denote the $b$th  sample by $\{y_t^{(b)}\}_{t=1}^N, b=1,\dots,B$.
  \item[2.] For each of the bootstrap time series $\{\mathbf{Y}_{t-1}^{(b)},y_t^{(b)}\}$, and given $p$ and $d$, estimate $\mathcal{S}_{\E[y_t\vert \mathbf{Y}_{t-1}]}$, and denote it by $\mathcal{\widehat{S}}^{(b)}({\sigma_{w,i}^{2}})$, where $i=1,\dots,\kappa$.
  \item[3.]  Now, calculate the distance between $\mathcal{\hat{S}}^{(b)}({\sigma_{w,i}^{2}})$ and $\mathcal{\hat{S}}({\sigma_{w,i}^{2}})$, and call it $D^{(b)}(\sigma_{w,i}^{2})$.
  \item[4.] Calculate the mean distance between {$\mathcal{\hat{S}}^{(b)}({\sigma_{w,i}^{2}})$ and $\mathcal{\hat{S}}({\sigma_{w,i}^{2}})$} over all   bootstrap samples   
\begin{equation}\label{eq:30}
  \bar{D}(\sigma_{w,i}^{2})=\frac{1}{B}\sum_{b=1}^B D^{(b)}(\sigma_{w,i}^{2}).
\end{equation}
\end{itemize}
The mean distance, $\bar{D}(\sigma_{w,i}^{2})$ is called the variability of $\mathcal{\hat{S}}({\sigma_{w,i}^{2}})$, and the estimator $\widehat{\sigma}_w^2$ is chosen to minimize the variability of $\mathcal{\hat{S}}({\sigma_{w,i}^{2}})$.
%%%%%%%%%%%%%%%%%%%%%%%%%%%%%%%%%%%%%%%%%%%%%%%%%%%%%%%%%%%%%%%%%%%%%%%%%%%%%%%%%%%%%%%%%%%%%%%%%%%%%%%%%%%%%%%%%%%
\subsection{\bfseries Algorithm for Estimating the TS-CMS}\label{Sec:5:3}
Assume the     lag order $p$ and the dimension of the TS-CMS ($d$) are known, then the $d$ leading eigenvectors of $\widehat{\mathbf{M}}_{FMT}$ form an orthogonal basis for $\mathcal{S}_{\E[y_t\vert \mathbf{Y}_{t-1}]}$. Thus, we use the first $d$ leading eigenvectors of $\widehat{\mathbf{M}}_{FMT}$ to construct a linear subspace and use it as an estimate of the TS-CMS and is denoted by $\widehat{\mathcal{S}}_{\E[y_t\vert \mathbf{Y}_{t-1}]}$. Since from Theorem 2, $\widehat{\mathbf{M}}_{FMT}$ converges to $\mathbf{M}_{FMT}$ at the rate of $\sqrt{N}$; therefore,  the eigenvectors and eigenvalues of $\widehat{\mathbf{M}}_{FMT}$ converge to those of $\mathbf{M}_{FMT}$ at the same rate.
%\textcolor[rgb]{1.00,0.00,0.00}{For real time series data, $\mathbf{Y}_{t-1}$ does not necessarily have zero mean and identity covariance matrix. However, standardization does not affect the convergence rate of the estimator, but it can increase the asymptotic variances (Zhu and Zeng, 2006).}
The estimation steps of the TS-CMS are as follows:
\begin{itemize}
	\item [1.] Estimate $p$,  $d$,  and  the tuning parameter $\sigma_w^2$  as described in Sections \ref{Sec:5:1} and \ref{Sec:5:2}. % , and choose the tuning parameter $\sigma_w^2$ as described in Section 4.2.
	%and $\tilde{y}_t=(y_t-\bar{y})$,
	\item [2.]  Center  $\mathbf{Y}_{t-1}$ %$\mathbf{Y}_{t-1}=(y_{t-1},\dots,y_{t-p})^T$
	 around the mean by subtracting the sample mean as
	$\widetilde{\mathbf{Y}}_{t-1}=\mathbf{Y}_{t-1}-\overline{\mathbf{Y}}_{t-1}$. 
	%and $\widehat{\boldsymbol{\Sigma}}$  is the sample covariance matrix of $\mathbf{Y}_{t-1}$.
	\item [3.] Calculate the candidate matrix $\widehat{\mathbf{M}}_{FMT}$ using $\{ \widetilde{\mathbf{Y}}_t,{y}_t\}_{t=p+1}^{N}$ as it is described in   (\ref{eq:23}).
	\item [4.] Obtain the eigenvector-eigenvalue pairs $(\hat{\mathbf{e}}_1,\hat{\lambda}_1),\dots,(\hat{\mathbf{e}}_p,\hat{\lambda}_p)$ of   $\widehat{\mathbf{M}}_{FMT}$ ({\small$\hat{\lambda}_1\geq,\dots,\geq\hat{\lambda}_p$}).
	\item [5.] Then, estimate the TS-CMS by calculating $\widehat{\mathcal{S}}_{\E[y_t\vert \mathbf{Y}_{t-1}]}=\textup{span}\{\hat{\mathbf{e}}_1,\dots,\hat{\mathbf{e}}_d \}$.	
\end{itemize}
Except that the candidate matrix $\mathbf{M}_{FMT}$ in step 3, which is estimated using  the Fourier method,  the rest follows the standard SDR procedure..  %the rest is the standard procedure in SDR method.  % sufficient dimension reduction. 
In the following section, we present numerical examples, with the R codes available from the authors upon request.
%In the following sections, we present the results of some numerical examples.  The R codes are available from the authors upon request.

\iffalse

In the final step, we obtain the TS-CMS by taking the product between the covariance matrix and the $d$ leading eigenvectors of $\widehat{\mathbf{M}}_{FMT}$. It would not make any difference if we use the standardized data or original data. This is because of the mathematical relationship between the TS-CMS obtained from the original data and that obtained from the standardized data. Let $\mathbf{\mathcal{Z}}_{t-1}=\boldsymbol\Sigma^{-\frac{1}{2}}(\mathbf{Y}_{t-1}-\overline{\mathbf{Y}}_{t-1})$ be the standardized data of $\mathbf{Y}_{t-1}$, where $\boldsymbol\Sigma=\Cov(\mathbf{Y}_{t-1})$. Then, it can be shown that (see, Cook and Li, 2002)
\begin{equation}\label{eq:31}
\mathcal{S}_{\E[y_t\vert\mathbf{Y}_{t-1}]}=\boldsymbol\Sigma^{-\frac{1}{2}}\mathcal{S}_{\E[y_t\vert\mathbf{\mathcal{Z}}_{t-1}]}.
\end{equation}
\fi 
 %%%%%%%%%%%%%%%%%%%%%%%%%%%%%%%%%%%%%%%%%%%%%%%%%%%%%%%%%%%%%%%%%%%%%%%%%%%%%%%%%%%%%%%%%%%%%%%%%%%%%%%%%%%%%%%%%%%%

\section{\large  Simulation Study}
\setlength{\abovedisplayskip}{6.pt}
\setlength{\belowdisplayskip}{6.pt}

In this section,  we conduct simulation studies to demonstrate  the FM approach's performance  %the performance of the FM approach
in estimating  TS-CMS and  TS-CVS and compare them with existing methods. 

First,  we introduce two distance measures that we use to measure the distance between two subspaces.   Let $\mathbf{A}$ and $\mathbf{B}$ be two full-rank matrices of size $p \times q$.   Suppose $\mathcal{S}(\mathbf{A})$ and $\mathcal{S}(\mathbf{B})$ are the column subspaces of $\mathbf{A}$ and $\mathbf{B}$, respectively. And, let $\lambda_i$'s with $1\geq \lambda_1^2 \geq,\dots,\lambda_p^2\geq 0$, are the eigenvalues of   $\mathbf{B}^T\mathbf{A}\mathbf{A}^T\mathbf{B}$.  The first measure is the vector correlation   defined as $\gamma=\sqrt{\frac{1}{p}\sum_{i=1}^{p}\lambda_i^2}$,  and the second one is the trace correlation given by  $	\rho= \sqrt{\prod_{i=1}^{p}\lambda_i^2}$. 
Both quantities $\gamma$ and  $\rho$      are in the range of 0 to 1. If $ \gamma=\rho=1$, it implies two spaces $\mathcal{S}(\mathbf{A})$ and $\mathcal{S}(\mathbf{B})$  are identical,  and if $\gamma=\rho=0$, then two spaces are perpendicular.  
 Moreover,  if  $d=1$ then it can be shown that  $ \gamma=\rho$. 
In order to better present and interpret the results, we report the first distance measure by D, where   $D=1-\gamma$. Then, the lower values of D   and the higher values of $\rho$  imply that the two spaces are closer.

 In the following simulation examples,  we compare our method with the only available method in the literature for estimating the  TS-CMS and TS-CVS  known as the Nadaray-Watson (NW) kernel smoother method (Park et al.,  2009; Park and Samadi, 2014).  To evaluate the robustness,  both normal and heavy-tailed distributions  ($t$ distribution) are considered for the innovations.   In the NW approach, we set the initial search to be $50$ before applying the sequential quadratic programming (SQP) algorithm to estimate the TS-CMS. Moreover, we conduct simulation studies with sample sizes of $N=50,100,300, 600,  800$,  with bootstrap sample size   $B=500$,  and the number of iterations ($500$).

\begin{exmp}\label{Model1}
(Nonlinear mean function) This time series model is Example 1 in both Xia and An (1999), and Park et al. (2009), and given by
\begin{equation*}\label{eq:26}	y_t=0.5\text{\{}cos(1.0)y_{t-1}-sin(1.0)y_{t-2}\}+0.4\exp\left\{-16\left[cos(1.0)y_{t-1}-sin(1.0)y_{t-2}\right]^2\right\}+0.1\varepsilon_t,
\end{equation*}
where the true subspace is $\boldsymbol\eta_1=(cos(1.0),-sin(1.0))^T$, $p=2$,  and $d=1$.  % and $\{\varepsilon_t\}$ is a sequence of  iid random variables with mean zero and variance one. 
\end{exmp}

Table S5 in Supplement C  %$\ref{tab:1}$ in \ref{Sup:B} 
shows the simulation results for identifying the optimal values of  $p$ and  $d$ using the (block) bootstrap mean distance $\bar{D}(p,q)$ criterion defined in   (\ref{eq:29}).  It shows that the optimal value of  ($p, d$)  is correctly identified and marked with a star.    A similar table is provided (Supplement C) %(in \ref{Sup:B})
 for all models to estimate $p$ and $d$.

%Table 1 and Figure 1 here  

%\vspace{-.3cm}
\begin{figure}[H]%[!ht]
\centering
%\begin{subfigure}{0.5\linewidth}
  \centering
  \includegraphics[width=0.45\textwidth, height=5cm]{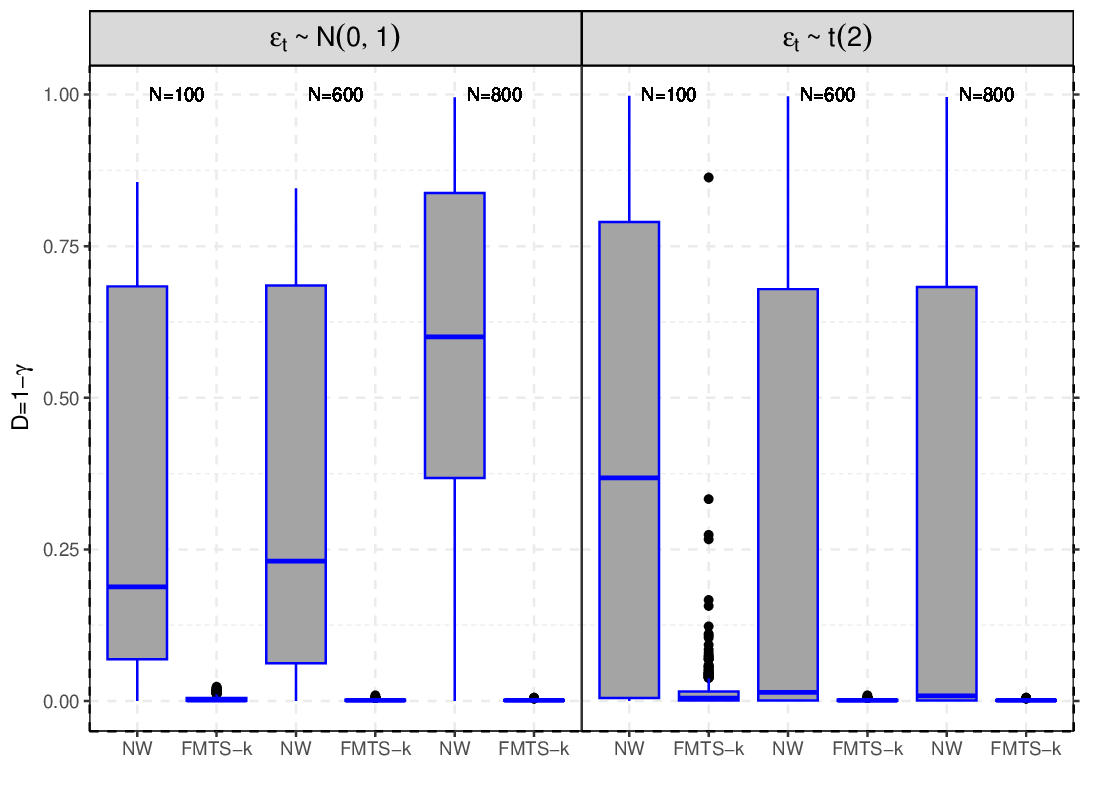}
  %\caption{$D=1-\gamma$}
  %\label{fig:sfig1}
%\end{subfigure}
%\begin{subfigure}{0.5\linewidth}
  \centering
  \includegraphics[width=0.45\textwidth, height=5cm]{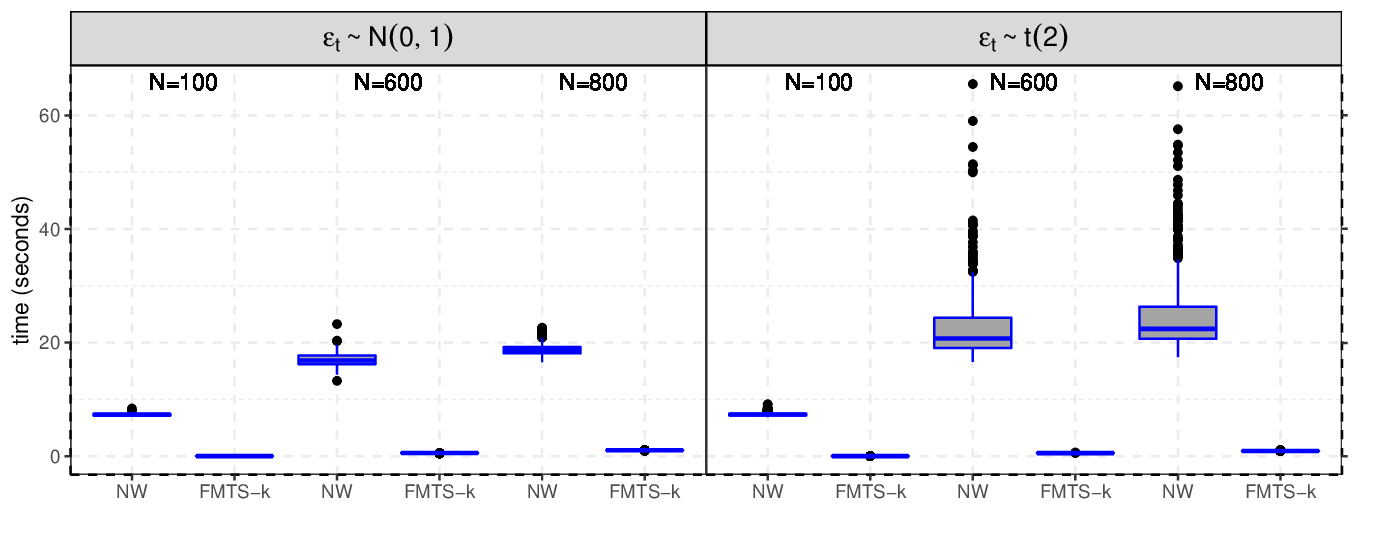}
  %\caption{Execution time per iteration}
  %\label{fig:sfig2}
%\end{subfigure}
\caption{\footnotesize  Simulation results for Model \ref{Model1}.  The left panel shows the estimation error of the distance between the estimated and the true TS-CMS for each method and different sample sizes under both normal and  $t$-distributions.  The right panel displays the corresponding execution time per iteration.  % The left part of each panel  is for the innovations are generated from a heavy tailed distribution
}
\label{model1R}
\end{figure}

%
%\begin{figure}[!ht]
%  \centering
%  \includegraphics[width=14cm,height=8cm]{Model1}
%  \caption{Mode 1 $D=1-\gamma$ and execution time per iteration of two methods when sample size is 600.}\label{model1R}
%\end{figure}

%########################################### model 2################
Figure \ref{model1R} and  Table  S9  in Supplement C % \ref{tab:2} (in \ref{Sup:B}) 
compare the performance of our proposed FMTS model with the NW method in estimating the TS-CMS. The results show that the FMTS method not only provides a more accurate and robust estimation than the NW approach but also reduces the computational time significantly. The column ``Time" shows the average computation time (in seconds) per sample.   It clearly demonstrates that our FMTS approach is significantly faster than the NW method.
%It shows that our FMTS approach is very efficient and computationally faster (10,000 times) than the NW method.   
As the sample size increases the precision of the estimate increases, and the estimation error shrinks towards zero.

%=================================================Model 2 =========================================================

%============================================Model 4 ===============================================================
\begin{exmp}\label{Model2}
(Nonlinear variance function)  This is an example from Park and Samadi (2014).  We estimate the TS-CVS of $\{x_t\}_{t=1}^N$, where
$\sqrt{x_t}$ is given by the following ARCH model
\begin{equation*}\label{eq:49odel4.1}
  \sqrt{x_t}=0.5\vert \varepsilon_t \vert \sqrt{1+(1/\sqrt{0.1^2+4^2})(0.1x_{t-1}+4x_{t-4})},
  %\sqrt{y_t}=0.5\vert \varepsilon_t \vert \sqrt{1+\frac{1}{\sqrt{2}}(\boldsymbol\Gamma_1^T\mathbf{X}_{t-1})+\cos\left\{0.1+\frac{1}{\sqrt{2}}(\boldsymbol\Gamma_2^T\mathbf{X}_{t-1})\right\}},
\end{equation*}
where $\mathbf{X}_{t-1}=(x^2_{t-1},\dots,x^2_{t-4})^T$, $q=4$, $d^{'}=1$, %$\varepsilon_t\overset{iid}{\sim} \mathcal{N}(0,1)$, 
and the true   subspace  is %of TS-CVS is
 $\boldsymbol\Gamma=(0.1,0,0,4)^T$.
\end{exmp}
 
% \ref{Tab:9} here 
 
From Figure \ref{model5R} below and Table S10 (in Supplement C),  % \ref{tab:10} (in \ref{Sup:B}), 
 they clearly show that the FMTS procedure for Model \ref{Model2} provides more robust and accurate results in contrast to the  NW approach. Moreover, the computational time of the FMTS approach is incomparably faster than the NW method, particularly for higher sample sizes.

%  Figure \ref{model5R}  here 
 
\vspace{-.2cm}
\begin{figure}[H]%[!ht]
\centering
%\begin{subfigure}{0.5\linewidth}
  \centering
  \includegraphics[width=6.95cm, height=5.cm]{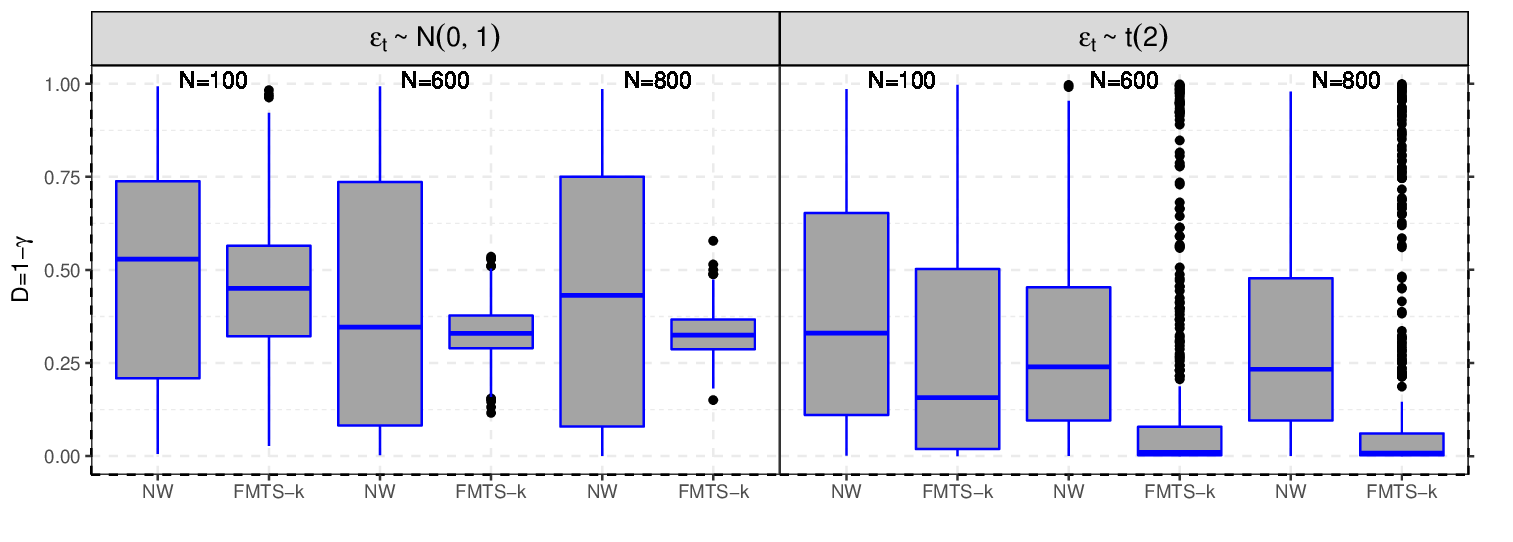}
  %\caption{$D=1-\gamma$}
  %\label{fig:sfig1}
%\end{subfigure}
%\begin{subfigure}{0.5\linewidth}
  \centering
  \includegraphics[width=6.95cm, height=5.cm]{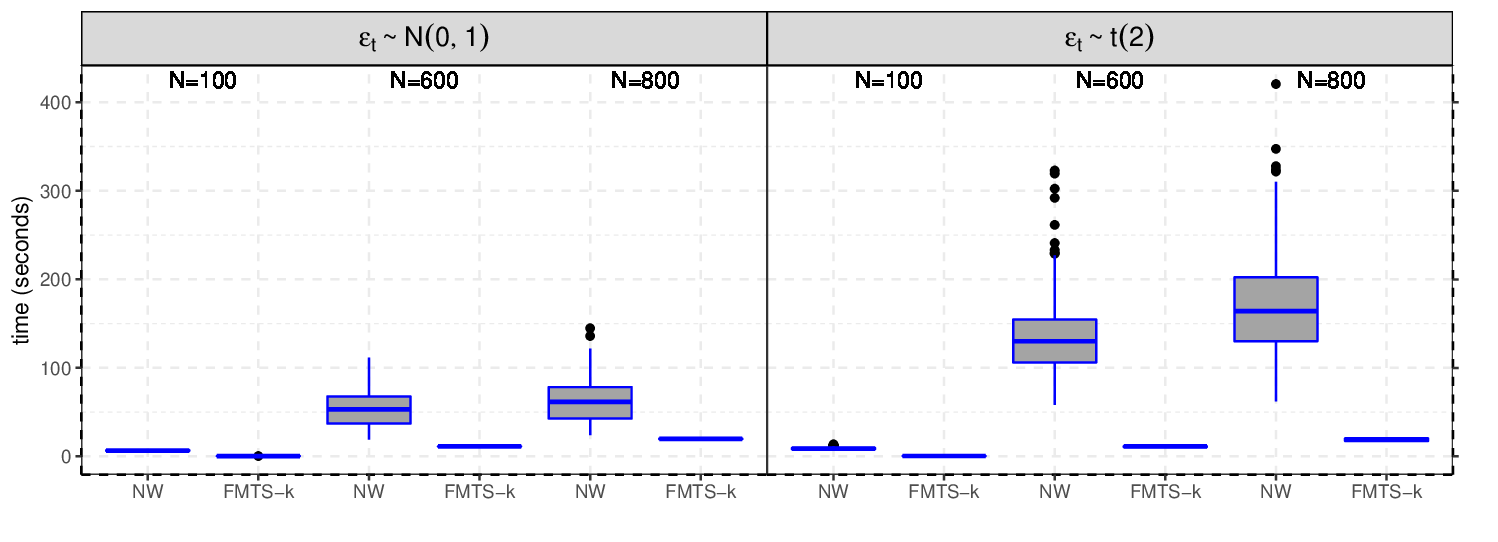}
  %\caption{Execution time per iteration}
  %\label{fig:sfig2}
%\end{subfigure}
\caption{ \footnotesize Simulation results for Model \ref{Model2}. The left panel shows the estimation error of the distance between the estimated and the true TS-CMS for each method and different sample sizes under both normal and  $t$-distributions.  The right panel displays the corresponding execution time per iteration. }
\label{model5R}
\end{figure}

%=========================== Model 5 +================================================================
\begin{exmp}\label{Model3}
This is a model with both mean and variance functions from Park and Samadi (2020).  Here we estimate the TS-CMS and the TS-CVS by using a two-step procedure.
\begin{align*}
  y_t&=3-\frac{1}{\sqrt{3}}(y_{t-2}+y_{t-4}+y_{t-6})+x_{t}, ~~~~~~~~~
  x_t=\varepsilon_t \sqrt{\frac{1}{\sqrt{6}}(2+x_{t-1}^2+x_{t-4}^2)},
\end{align*}
where $p=6$, $q=4$, $d=d^{'}=1$, %and $\varepsilon_t \overset{iid}{\sim} \mathcal{N}(0,1)$, 
and the true column {subspaces} of the TS-CMS and the TS-CVS are $\boldsymbol\eta_1=(0,1,0,1,0,1)^T$, $\boldsymbol\Gamma_1=(0,1,0,1)^T$, respectively.
\end{exmp}

% Tables \ref{tab:11} and \ref{tab:12} here 

From Figure \ref{model6M} below and Table S11 (in Supplement C),  %\ref{tab:13} (in \ref{Sup:B}),  
it can be seen that like all other examples, the FMTS method substantially outperforms the NW method in terms of both estimation accuracy and computational efficiency.  It shows a better estimation and time efficiency in estimating both the TS-CMS and the TS-CVS for all sample sizes. 

To conclude this section, from three examples here and two additional examples in  Supplement B,   % \ref{Sup:A}, 
 we conclude that our proposed FMTS method not only performs significantly better than the NW method in estimating the TS-CMS and the TS-CVS but also computationally substantially (approximately 10,000 times) faster than the NW method. In particular, the  FMTS method is more accurate and efficient when the errors are heavy-tailed.

% Figure \ref{model6M} here 

%\vspace{-.2cm}
 \begin{figure}[H]%[!htbp]
 %\vspace{-1.cm}
   \centering
 \begin{subfigure}[b]{0.5\textwidth}
                \includegraphics[width=\textwidth,height=5.cm]{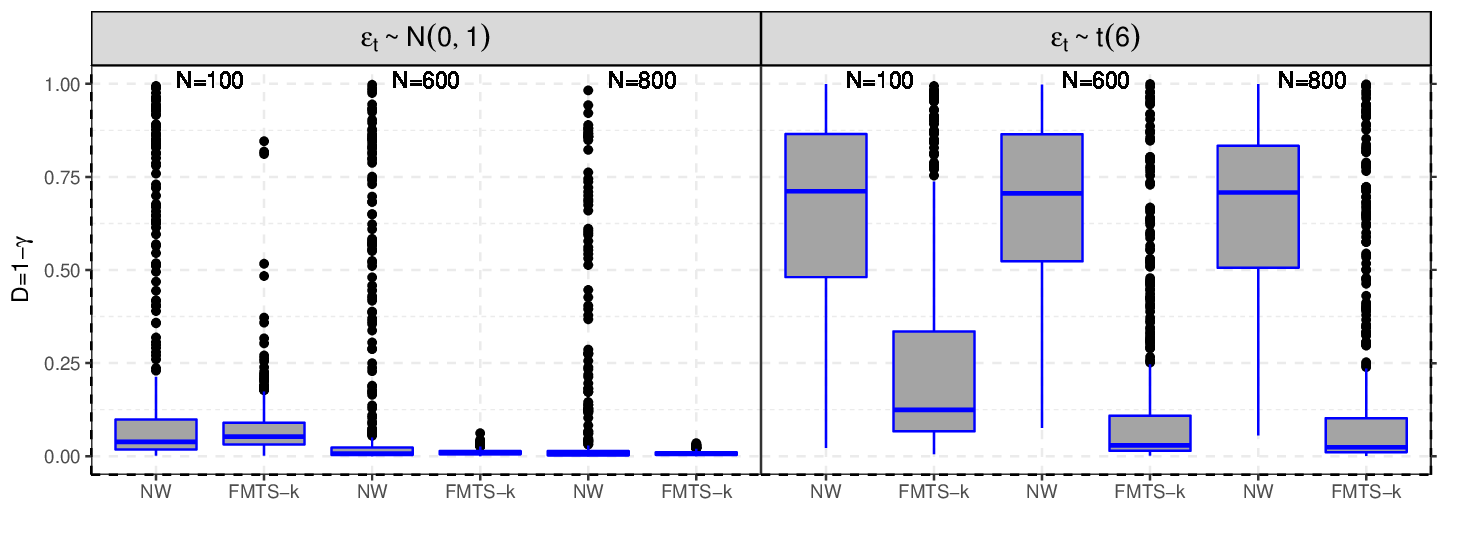}
               % \caption{}
        \end{subfigure}%
         \hspace{-.4cm}
        %add desired spacing between images, e. g. ~, \quad, \qquad etc.
           \begin{subfigure}[b]{0.5\textwidth}
                \includegraphics[width=\textwidth,height=5.cm]{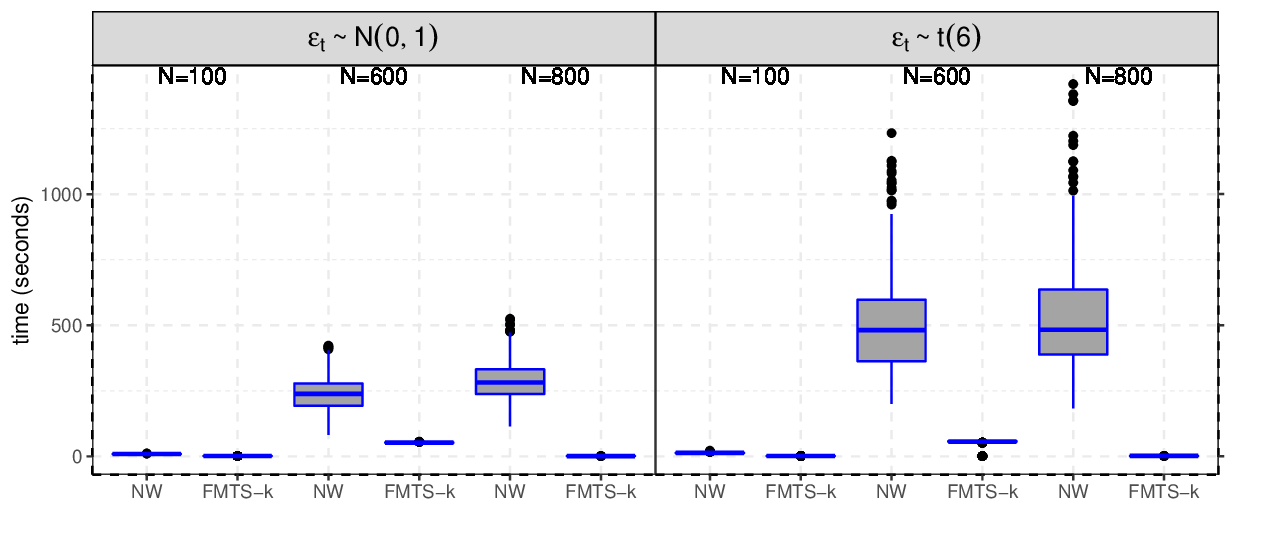}
               % \caption{}
        \end{subfigure}
  %  \vspace{-.4cm}
        \centering
 \begin{subfigure}[b]{0.5\textwidth}
                \includegraphics[width=\textwidth,height=5.cm]{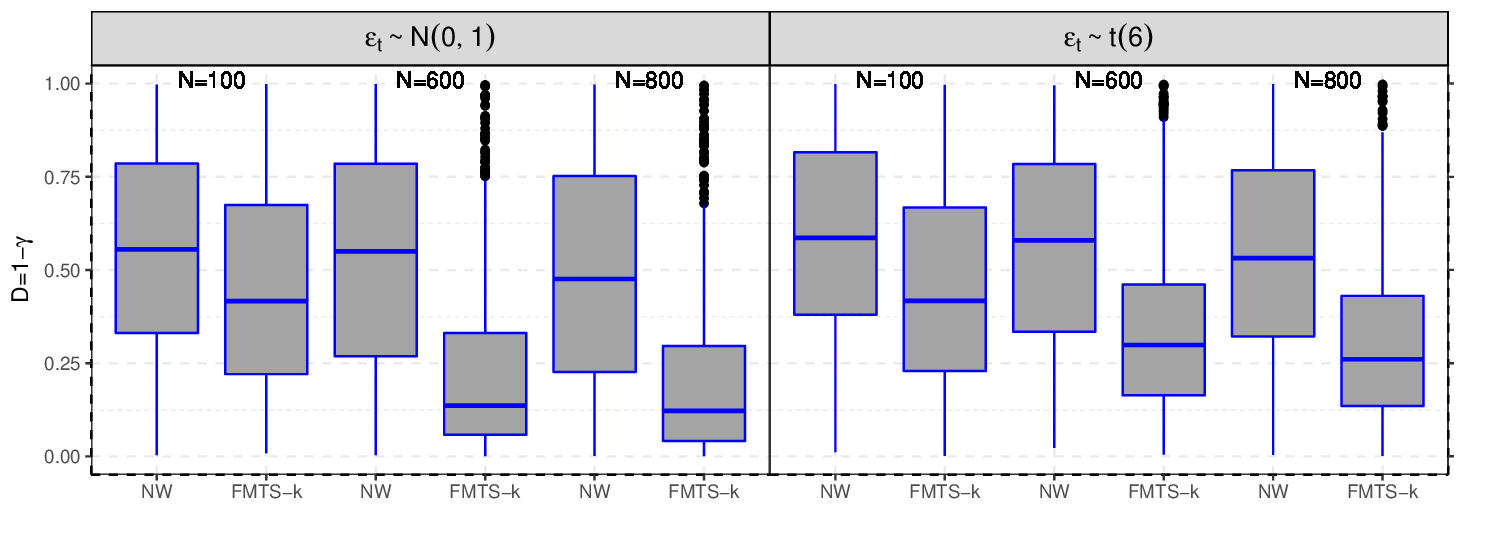}
             %   \caption{}
        \end{subfigure}%
      \hspace{-.3cm}  %add desired spacing between images, e. g. ~, \quad, \qquad etc.
           \begin{subfigure}[b]{0.5\textwidth}
                \includegraphics[width=\textwidth,height=5.cm]{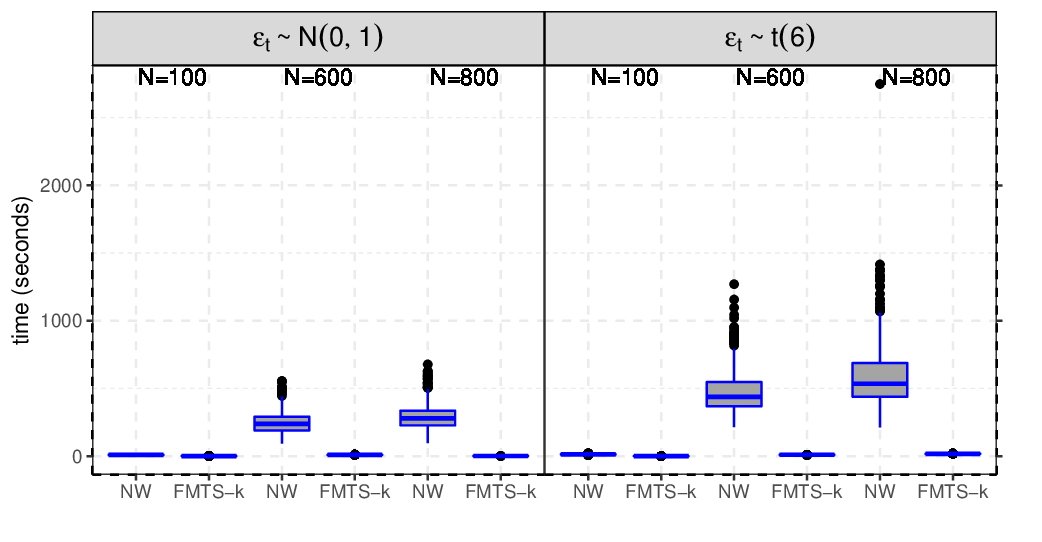}
             %   \caption{}
        \end{subfigure}
 % \label{fig2}
                %  \caption{}
    %\end{figure}
    % \begin{figure}
    
  \caption{ \footnotesize   Model \ref{Model3} Results:    %TS-CMS is  in the top row, and the TS-CVS is in the  bottom row.  
   The left panels show estimation errors of the distance between the estimated and the true TS-CMS (top) and TS-CVS (bottom) for each method and different sample sizes under both normal and  $t$-distributions. The right panels display the corresponding average execution time per iteration.  }
\label{model6M}
  \end{figure}

 \section{\large  Empirical Application}
\setlength{\abovedisplayskip}{6.pt}
\setlength{\belowdisplayskip}{6.pt}
 
Several newly developed time series methods have been applied to analyze the {\it Canadian Lynx Data} and compared their performances with existing methods. The {\it Canadian Lynx Data} is about predator and prey interactions, food web, etc,  (Elton and Nicholson, 1942).  %Moreover, the NW method has been used to analyze this data  by Park et al. (2009). Therefore, 
We are motivated to implement our proposed FMTS method to analyze the {\it Canadian Lynx Data}, and compare the results with that of previous works. This dataset consists of the annual record of the number of the Canadian Lynx ``trapped'' in the Mackenzie River district of   North-West Canada for the period 1821–1934 ({the dataset available at:} {https://www.encyclopediaofmath.org /index.php/Canadian\_lynx\_data}).

This dataset has been studied by several authors, including Moran (1953) who at the first time fitted an AR(2) model to the log of the number of Lynx, i.e.,  $y_{t}=\log_{10}(\text{number of lynx  }$\\ $\text{trapped in year 1820 + t}),$ where $t=1,\dots,114 $. Tong (1990) proposed the class of self-exciting threshold autoregressive model, abbreviated as $SETAR(k;d_1,d_2,\dots,d_k)$, where $k$ is the number of thresholds, and $d_i$s are the number of delays in each threshold, to analyze the Lynx dataset. His proposed models are SETAR(2;2,2) and SETAR (2;7,2) (Tong, 1990, p.377, (7.3) and p.386, (7.6), respectively). Moreover, Tsay (1988) developed a SETAR (3;1,7,2) model to analyze the Lynx dataset (Tong 1990, p.405, (7.8)). We compare the performance of our model with the NW model proposed by Park et al. (2009) and three other competitive models proposed by Tong and Tsay. The performances of the models are measured by the mean absolute relative error (MARE), the mean squared error (MSE), and the mean square relative error (MSRE), respectively,  given as  $MARE=\frac{1}{N-p}\sum_{t=p+1}^{N}\{\vert y_t-\hat{y}_t \vert/y_t \}$,    $MSRE=\frac{1}{ N-p}\sum_{t=p+1}^{N}\{\left(y_t-\hat{y}_t \right)^2/y_t \}$, and    $MSE=\frac{1}{ N-p}\sum_{t=p+1}^{N}(y_t-\hat{y}_t)^2$;
%\begin{equation}\label{eq:34}
%  \begin{aligned}
%   MARE&=\frac{1}{N-p}\sum_{t=p+1}^{N}\{\vert y_t-\hat{y}_t \vert/y_t \}, ~~~~~~
%   MSRE=\frac{1}{ N-p}\sum_{t=p+1}^{N}\{\left(y_t-\hat{y}_t \right)^2/y_t \},\\
%   MSE&=\frac{1}{ N-p}\sum_{t=p+1}^{N}(y_t-\hat{y}_t)^2,
%  \end{aligned}
%\end{equation}
where $N$ is  sample size and $p$ is the lag order.  Smaller values of  them % MARE and MSRE 
indicate a better fit.

Park et al. (2009) fitted the following model to the $\log_{10}$ of the {\it Canadian Lynx Data},
\begin{equation}\label{eq:35}
  \widehat{y}_t=0.99+0.52y_{t-1}+0.75d_{1,t}-0.39d_{1,t-1}-0.13cos_t+0.07cos_{1,t-1}%+\varepsilon_t,
\end{equation}
where   $d_{1,t}=\widehat{\boldsymbol\phi}^T_1\mathbf{Y}_{t-1}$ with the   basis  vector   $\widehat{\boldsymbol\phi}_1=(0.9317,-0.0761,-0.1777,-0.3074)^T$  of  the TS-CMS derived by  the NW method,  $cos_t=\cos(3.87d_{1,t}-3.44)$,    and $\mathbf{Y}_{t-1}=(y_{t-1}, \dots, y_{t-4})^T$.
Tong (1990) fitted the following SETAR(2;2,2) model to the  series  %$\log_{10}$ of the {\it Canadian Lynx Data}
\begin{equation}\label{eq:36}
	\begin{aligned}
  \widehat{y}_t=& \{0.62+1.25y_{t-1}-0.43y_{t-2}+\varepsilon_t^{(1)}\}{I}(y_{t-2}\leq 3.25)\\
  &~~~+\{2.25+1.52y_{t-1}-1.24y_{t-2}+\varepsilon_t^{(2)}\}{I}(y_{t-2}> 3.25),
	\end{aligned}
\end{equation}
where $\{\varepsilon_t^{(1)}\}$, and $\{\varepsilon_t^{(2)}\}$  are white noise sequences independent of $(y_{t-1},y_{t-1})^T$ and independent of each other. By choosing the model with the lowest Akaike's information criterion (AIC) value, Tong (1990) proposed the SETAR(2;7,2) model given in   (\ref{eq:37}) to fit the same dataset. This is an extension of the model given in   (\ref{eq:36}), and given as
\begin{equation}\label{eq:37}
\begin{aligned}
	\widehat{y}_t&=\{0.546+1.032y_{t-1}-0.173y_{t-2}+0.171y_{t-3}-0.43y_{t-4}+0.332y_{t-5}\\
    &~~~~~ -0.284y_{t-6}+0.210y_{t-7}+\varepsilon_t^{(1)}\}{I}(y_{t-2}\leq+ 3.116)
	\\& ~~~~~~ +\{2.632+1.492y_{t-1}-1.324y_{t-2}+\varepsilon_t^{(2)}\}{I}(y_{t-2}> 3.116).
\end{aligned}
\end{equation}

A nonlinear model similar to Tong's models in   (\ref{eq:36}) and (\ref{eq:37}) developed by Tsay (1988),  % to analysis the {\it Lynx Data}. 
where  he fitted an  SETAR(3;1,7,2) model (with three thresholds) to the series as  %fit the $\log_{10}$ of the Lynx dataset, give as
\small
\begin{align}\label{eq:38}
    \begin{split}
  \widehat{y}_t =& \{0.083+1.096y_{t-1}+\varepsilon_t^{(1)}\}{I}(y_{t-2}\leq 2.373)\\
  & ~~~ +\{0.63+0.96y_{t-1}-0.11y_{t-2}+0.23y_{t-3}-0.61y_{t-4}+0.48y_{t-5}-0.39y_{t-6}\\
  &~~~+0.28y_{y-7}+\varepsilon_t^{(2)}\}{I}( 2.373<y_{t-2}\leq 3.154)\\
  &~~~ +\{2.323+1.530y_{t-1}-1.266y_{t-2}+\varepsilon_t^{(3)}\}{I}(y_{t-2}>3.154), 
    \end{split}
\end{align}
\normalsize
where $\{\varepsilon_t^{(i)}\}$, $i=1,2,3,$ are white noises 
 independent of $(y_{t-1},\dots,y_{t-7})$,  and  of each other. 

\begin{figure}[H] 
  \centering
  \includegraphics[width=11cm, height=5.2cm]{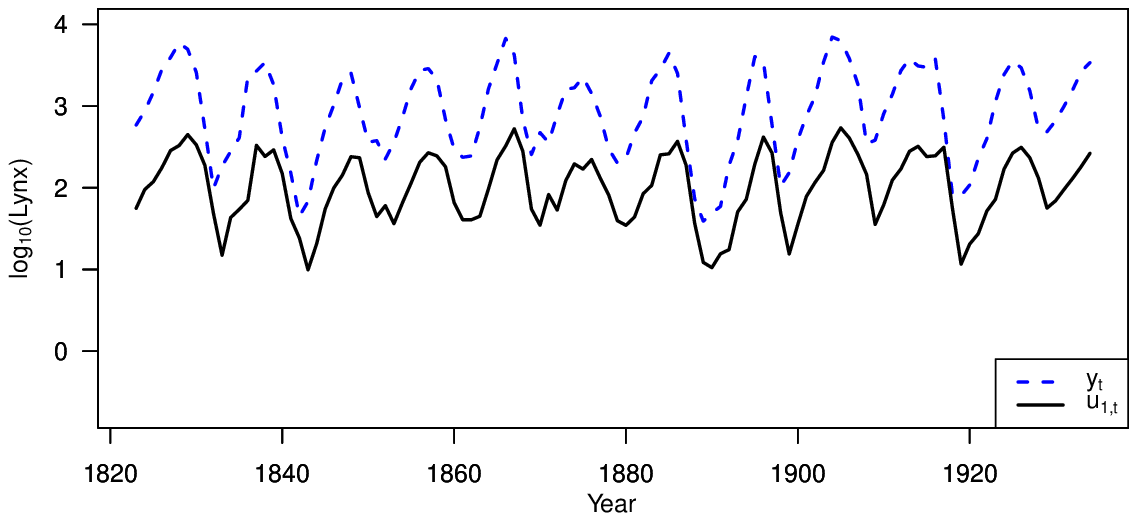}
  \caption{Time series plot of $y_t$ and $u_{1,t}=\widehat{\boldsymbol\eta}_1^T\mathbf{Y}_{t-1}$ of the $\log_{10}$ of  the  {\it Canadian Lynx dataset}. } \label{fig:5.1.1}
\end{figure}

Now, we use our proposed  Fourier method to estimate the TS-CMS of {\it Canadian Lynx} time series dataset. As the initial step,  we estimate the lag order $p$ and the dimension $d$ of the TS-CMS following the procedures described in Section \ref{Sec:5:1}.    Table \ref{tab:14} %(\ref{tab:14})
shows the variability measure  $\bar{D}(p,d)$ for all candidate pairs of (p,d) when ${p=2,\dots,7}$,  with the optimal value being   $(\hat{p}=2,\hat{d}=1)$. The next step is to estimate the tuning parameter $\sigma_w^2$  using the method discussed in Section  \ref{Sec:5:2}. The minimum value of the variability measure $\bar{D}(\sigma^2_{w,i})$ in   \eqref{eq:30} is achieved when $\widehat{\sigma}_w^2=0.01$.  % {among the candidates $\{0,0.05,\dots,0.45,0.5\}$}. 
Since $\widehat{d}=1$, the final step is to estimate $\widehat{\boldsymbol\eta}_1$.   Following the estimation procedure discussed  in Section \ref{Sec:5:3}, the single index direction is obtained  as {$\widehat{\boldsymbol\eta}_1=(0.9621,  -0.2727)^T$}.  %We follow a similar procedure as in Park et al. (2009) to fit a     model to the $\log_{10}$ of the {\it  Lynx Data}. 
Then, we display the time series plot of $y_t$ and $u_{1,t}=\widehat{\boldsymbol\eta}_1^T\mathbf{Y}_{t-1}$ versus $t$ in Figure \ref{fig:5.1.1},  where $\mathbf{Y}_{t-1}=(y_{t-1},y_{t-2})^T$.  According to this plot, $u_{1,t} $ adequately captures the dynamic fluctuations and periodic patterns of the entire time series $y_t$.
 
 % Table \ref{tab:14} and Figure \ref{fig:5.1.1} here 

 \begin{table}[!ht]
	\centering
\caption{Mean distance for pairs (p,d).in   Canadian Lynx data.  ${\star}$ denotes the optimal value of $(p,d)$} \label{tab:14}  % i.e., $(\hat{p},\hat{d})$=(2,1)
	\begin{tabular}{|c|c|c|c|c|c|c|c|c|c|}
		\hline
		p & d=1 & d=2 & d=3 & d=4& d=5&d=6\\ \hline
2& 0.0003 ${\star}$& 0& & &&    \\
3& 0.0199& 0.0058& 0&&&    \\
4& 0.0121& 0.1137& 0.0143 &0&&    \\
5& 0.0343& 0.1665& 0.1347 &0.0078 &0&   \\
6& 0.0398& 0.1453& 0.0921 &0.0465 &0.0492 &0      \\
7& 0.0329& 0.2043& 0.1324 &0.0897 &0.0849 &0.0154     \\
%8& 0.0724& 0.1353& 0.1674 &0.1386 &0.1017 &0.0689 & .0522  \\
%9& 0.1134& 0.2011& 0.1582 &0.1281 &0.1303 &0.1012 &0.0847 &0.0413 &0     \\
%10& 0.0927& 0.1811& 0.1822 &0.1560 &0.1357 &0.0987 &0.0781 &0.0722 &0.0357     \\
		\hline
	\end{tabular}	
\end{table}

\vspace{-.2cm}

%%%%%%%%%%%%%%%%%%%%%%%%%%%

Because of the linear relationship between $y_t$ and  $ \widehat{\boldsymbol\eta}_1^T\mathbf{Y}_{t-1}$,  we regressed $y_t$ on $ u_{1,t}= \widehat{\boldsymbol\eta}_1^T\mathbf{Y}_{t-1}$ which  gives the following linear regression model % (with $R^2= 0.7545.$)
\begin{equation*}%\label{eq:38.22}
 % y_t=\beta_1 u_t+\varsigma_t,
  \widehat{y}_t= 0.62202 + 1.14294 u_{1,t}.
\end{equation*}
 %where the  coefficients are significant with standard errors $0.12549$ and $0.06128$.
 However, after removing the mean function,   we plot the residuals versus $u_{1,t}$  which shows some nonlinear dependencies with $u_{1,t}$.     In order to capture the unexplained nonlinear structure,  we remodel the data by adding a nonlinear component in the previous model.  After several trials and errors,  we fit the data with the following nonlinear model  (with $R^2= 0.7545$)
 \begin{equation}\label{eq:38.23}
  \widehat{y}_t=1.7770 u_{1,t} -0.0809 \exp (u_{1,t}). 
\end{equation}
  The residuals of this model ($\eta_t$) show some cyclic behavior with a period of ten. Hence, we fit the residuals using a seasonal autoregressive process of order one,  i.e.,   $SAR(1)_{10}$ given as $\eta_t=0.6209 \eta_{t-10}+\varepsilon_t$.   This leads  to the following fitted  model (with $R^2= 0.8558$) 
    \begin{equation}\label{eq:38.24}
    \widehat{y}_t=1.7770 u_{1,t}-0.0809 \exp(u_{1,t})+0.6209 y_{t-10}-1.1033 u_{1,t-10}+0.0502 \exp(u_{1,t-10}) 
  \end{equation}
 
 Although the proposed   model \eqref{eq:38.24} fits  the data well,   to  further explore  the unexplained variations in the residuals,  we also   fit the residuals of model \eqref{eq:38.23}  using  a seasonal autoregressive  process of order two ($SAR(2)_{10}$) as  $\eta_t=0.4764\eta_{t-10}+0.2414\eta_{t-20} + \varepsilon_t$, which gives the following fitted model (with $R^2=  0.8659$) 
 
 \vspace{-.1in}
 \begin{align}\label{eq:38.25}
  \begin{split}
    y_t&=1.777u_{1,t}-0.0809 \exp(u_{1,t})+0.4768y_{t-10}-0.8466 u_{1,t-10}+0.0385 \exp(u_{1,t-10})\\
    &~~~~~~~+0.2414 y_{t-20}-0.4290 u_{1,t-20}+0.0195 \exp(u_{1,t-20})
  \end{split}
  \end{align}

 \vspace{-.1in}

% Table \ref{tab:21}  here 
  
\begin{table}[h!]
	\caption{Performance comparison:   FMTS model vs. other time series models on  the Lynx data}\label{tab:21}
	\centering
\begin{tabular}{|c|c|c|c|c|c|}
	\hline
	Model Equation & MARE & MSRE & MSE& n & no. of parameters\\  \hline
    FMTS-k (\ref{eq:38.24})& 0.06127 & 0.01735& 0.04671 &110& 6 \\
    FMTS-k (\ref{eq:38.25})&\bf{0.05856} & \bf{0.01624}& \bf{ 0.04403} &110& \bf{7} \\
	%FMTS-(\ref{eq:40})&   0.06828 &0.01993 &107 &5\\
	NW (\ref{eq:35}) &  0.07317 &0.02483 &0.06135 &110& 10\\
	Tong's (\ref{eq:36})&0.05826 &0.01581 &0.04192 &110 &8\\
	Tong's (\ref{eq:37}) &0.05488 &0.01334 &0.03584 &110 &13\\
	Tsay's (\ref{eq:38}) &0.05421 &0.01264 &0.03424 &110 &17 \\
	\hline
\end{tabular}
\end{table}
%%%%%%%%%%%%%%%%%%5

We compare the performance of our model in   (\ref{eq:38.24}) and (\ref{eq:38.25}) with the NW model given in   (\ref{eq:35}), and three SETAR models (Tong's models,  and Tsay model). Table   \ref{tab:21} shows the MARE, MSRE, and MSE values, and the total number of estimated parameters in each model.

\vspace{-.1cm}
\begin{figure}[H]%[h!]
  \centering
       \includegraphics[width=12cm,height=5.8cm]{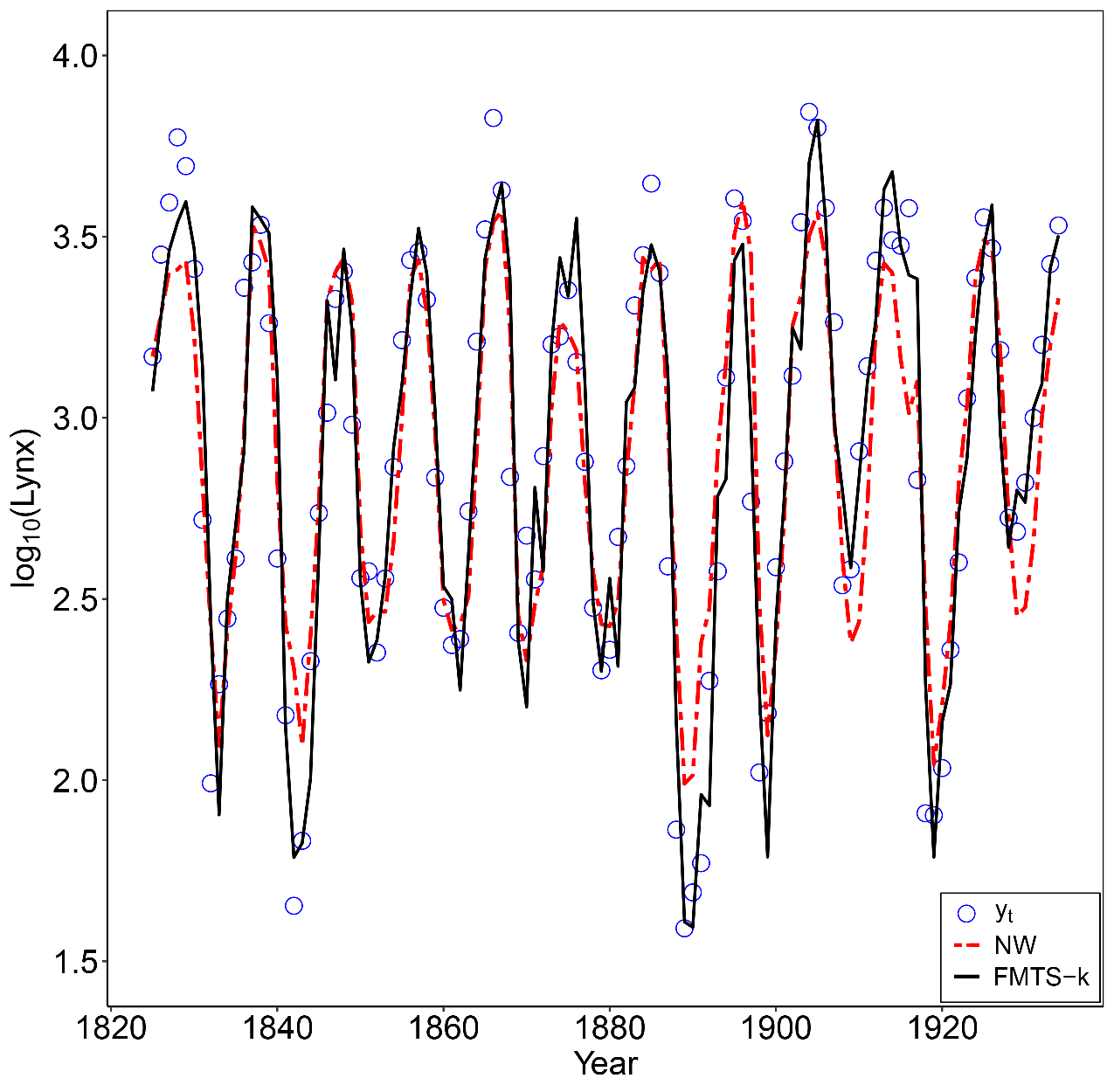}
     \caption{Observed and fitted values of the annual Canadian  Lynx data,  1821-1930. }\label{fig:6.1}
\end{figure} 

{Compared to the NW approach, our proposed model has significantly better accuracy with a fewer number of parameters. Moreover, our model uses sufficient information in the initial modelling part as the NW method does. On the other hand, our model is more parsimonious compared to the SETAR models}. Figure \ref{fig:6.1} shows the observed Lynx series from 1821 to 1930 and the fitted time series obtained from the NW model and our FMTS proposed model. The open blue circles are the observed Lynx series, and the dashed red line and the solid black line represent its fitted series based on the NW and our FMTS method, respectively. Moreover, in comparison to the NW method,  the fitted values show that our estimator smooths out extreme points better.

% Figure \ref{fig:6.1} here 

\section{\large  Conclusion and Discussion}
\vspace{-.1in}
In this paper, we proposed and developed a new sufficient dimension reduction approach for time series by using the Fourier transform technique to estimate the time series central mean and variance subspaces (TS-CMS and TS-CVS).    By Fourier transform,  we derived  candidate matrix   ${\mathbf{M}}_{FMT}$  (${\mathbf{M}}_{FVT}$) to extract the TS-CMS  (TS-CVS).   Under some distributional assumptions,   we provided an explicit expression to estimate the candidate matrices that lead to estimates of the mean and variance subspaces. 
We established the asymptotic consistency of the proposed estimator under normality and under the more general assumptions,  and the convergence rate of the proposed estimator is obtained.   Moreover, we have developed a data-driven method to estimate the unknown lag order $p$ and the dimension of the TS-CMS   (TS-CVS).    Simulation results and real data analysis show that our proposed method is significantly more efficient and faster than the existing methods in the literature to estimate the TS-CMS and TS-CVS.  % proposed by Park et al. (2009).

\section*{\large  Appendix:  Proofs of Propositions and Theorems}

\setlength{\abovedisplayskip}{6.pt}
\setlength{\belowdisplayskip}{6.pt}

%\begin{proof}{ \mathbf{Proof of Theorem 1:} }

 \textbf{Proof of Theorem 1:}
{ The proof is similar to the proof of Equation (5) in Zhu and Zeng (2006). Let  $\boldsymbol\eta=(\boldsymbol\eta_1,\dots,\boldsymbol\eta_d) \in \mathbb{R}^{p \times q}$, then the columns of $\boldsymbol\eta$, i.e., $\boldsymbol\eta_1,\dots,\boldsymbol\eta_d$ form a basis for $\mathcal{S}_{\E[y_t\vert \mathbf{Y}_{t-1}]}$ with dimension $d$. To proof Theorem \ref{th:1} it is enough to show that for any $\boldsymbol\beta \in \mathbb{R}^p$, the condition $\boldsymbol\beta^T \boldsymbol\eta=0$ is same as $\boldsymbol\beta^T \frac{\partial}{\partial\mathbf{y}_{t-1}}m(\mathbf{y}_{t-1})=0$,  for all $\mathbf{y}_{t-1}\in$ supp($\mathbf{Y}_{t-1}$).
Using the chain rule of differentiation we have $\frac{\partial}{\partial \mathbf{y}_{t-1}}m(\mathbf{y}_{t-1})=\boldsymbol\eta \frac{\partial}{\partial \mathbf{u}_{t-1}} g(\mathbf{u}_{t-1})$. Therefore, $\boldsymbol\beta^T \boldsymbol\eta=0$ implies that $\boldsymbol\beta^T\frac{\partial}{\partial \mathbf{y}_{t-1}} m(\mathbf{y}_{t-1})=\boldsymbol\beta^T\boldsymbol\eta\frac{\partial}{\partial \mathbf{u}_{t-1}}g(\mathbf{u}_{t-1})=0$,  for all $\mathbf{y}_{t-1} \in $ supp($\mathbf{Y}_{t-1}$).}

{
The backward direction can be proven by contradiction. Suppose that there exists a $\boldsymbol\beta_0 \in \mathbb{R}^p$ such that $\boldsymbol\beta_0^T\frac{\partial}{\partial \mathbf{y}_{t-1}}m(\mathbf{y}_{t-1})=0$ for all $\mathbf{y}_{t-1} \in $ supp($\mathbf{Y}_{t-1})$, but $\boldsymbol\beta_0^T\boldsymbol\eta \neq 0$. Let $\boldsymbol\zeta_1=\boldsymbol\eta^T\boldsymbol\beta_0/\vert \vert \boldsymbol\eta^T\boldsymbol\beta_0\vert\vert$. Clearly, $\boldsymbol\zeta_1$ is a $d$-dimensional non-zero vector. So if $\boldsymbol \beta_0^T\frac{\partial}{\partial \mathbf{y}_{t-1}}m(\mathbf{y}_{t-1})=\boldsymbol\beta_0^T\boldsymbol\eta\frac{\partial}{\partial\mathbf{u}}g(\mathbf{u})=0$, then $\boldsymbol\zeta_1^T\frac{\partial}{\partial \mathbf{u}}g(\mathbf{u})=0$, which implies the directional derivative of $g(\mathbf{u})$ along $\boldsymbol\zeta_1$ is always 0, where $\mathbf{u}=({u}_1,\dots,{u}_d)^T$. Moreover, this implies that $g(\mathbf{u})$ is a constant along $\boldsymbol \zeta_1$, i.e., $g(\mathbf{u}+\alpha\boldsymbol\zeta_1)=g(\mathbf{u})$ for $\alpha \in \mathbb{R}$. We can construct an orthogonal basis for $\mathbb{R}^d$ by expanding $\boldsymbol\zeta_1$ and then using  $\boldsymbol\zeta_1,\boldsymbol\zeta_2,\dots,\boldsymbol\zeta_d$. Let $\mathbf{D}=(\boldsymbol\zeta_1,\dots,\boldsymbol\zeta_d)$ and $\mathbf{v}=\mathbf{D}^T\mathbf{u}=(v_1,\dots,v_d)^T$; then  $g(\mathbf{u})=g(\mathbf{D}\mathbf{v})$ and since $\frac{\partial}{\partial v_1}g(\mathbf{D}\mathbf{v})=\boldsymbol\zeta_1^T\frac{\partial}{\partial\mathbf{u}}g(\mathbf{u})=0$; hence $g(\cdot)$ does not depend on $v_1$. Therefore, we can rewrite $g(\mathbf{u})=g(\mathbf{D}\mathbf{v})=\tilde{g}(v_2,\dots,v_d)=\tilde{g}(\boldsymbol\zeta_2^T\boldsymbol\eta\mathbf{y}_{t-1},\dots,\boldsymbol\zeta_d^T\boldsymbol\eta\mathbf{y}_{t-1})$, which implies that $\mathcal{S}(\boldsymbol\eta\boldsymbol\zeta_2,\dots,\boldsymbol\eta\boldsymbol\zeta_d)$ is also a dimension-reduction subspace for $\E[y_t\vert \mathbf{Y}_{t-1}]$ and that the central mean subspace has dimension at most $d-1$, which is contradicted to $dim(\mathcal{S}_{\E[y_t\vert \mathbf{Y}_{t-1}]})=d$. Thus,  the proof is completed.   
 $~~~~~~~~~~~~~~~~~~~~~~~~~~~~~~~~~~~~~~~~~~~~~~~~~~\square$ } \\

\textbf{Proof of Proposition \ref{prop:2.2}:}
Since $\boldsymbol\psi_t(\boldsymbol\omega)=\mathbf{a}_t(\boldsymbol\omega)+i\mathbf{b}_t(\boldsymbol\omega)$, we have,
\begin{align}\label{eq:42}
%\begin{split}
  \boldsymbol\psi_t(\boldsymbol\omega)\overline{\boldsymbol\psi_s}^T(\boldsymbol\omega)= [\mathbf{a}_t(\boldsymbol\omega)\mathbf{a}_s^T(\boldsymbol\omega)+\mathbf{b}_t(\boldsymbol\omega)\mathbf{b}_s^T(\boldsymbol\omega)]
  +i\left [\mathbf{b}_t(\boldsymbol\omega)\mathbf{a}_s^T(\boldsymbol\omega)-\mathbf{a}_t(\boldsymbol\omega)\mathbf{b}_s^T(\boldsymbol\omega)\right].
%\end{split}
\end{align}
By   (\ref{eq:13}), it easily can be seen that the imaginary part is not available in the left hand side of    (\ref{eq:13}),  thus $\int\left[ \mathbf{b}_t(\boldsymbol\omega)\mathbf{a}_s^T(\boldsymbol\omega)-\mathbf{a}_t(\boldsymbol\omega)\mathbf{b}_s^T(\boldsymbol\omega)\right] d\boldsymbol\omega=0$. Therefore,
\begin{align}\label{eq:43}
  \mathbf{M}^*_{FMT} =&(2\pi)^{-p} \int \boldsymbol\psi_t(\boldsymbol\omega)\bar{\boldsymbol\psi}_s^T(\boldsymbol\omega)d\boldsymbol\omega
  = (2\pi)^{-p} \int [\mathbf{a}_t(\boldsymbol\omega)\mathbf{a}_s^T(\boldsymbol\omega)+\mathbf{b}_t(\boldsymbol\omega)\mathbf{b}_s^T(\boldsymbol\omega)]d\boldsymbol\omega.
\end{align}

Let $\mathbf{U}=\textup{span}\{\mathbf{a}_1,\dots,\mathbf{a}_m\}$ with $\mathbf{a}_i \in \mathbb{R}^p$, and let $\mathbf{U}^{\perp}$ be the orthogonal complement of $\mathbf{U}$ where $\mathbf{U}^{\perp}=\{\mathbf{w}\in \mathbb{R}^p; \mathbf{w}^T\mathbf{a}_1\mathbf{w}=\dots =\mathbf{w}^T\mathbf{a}_m\mathbf{w}=0\}$. Define a positive definite matrix as $ \mathbf{M}^*=\sum_{i=1}^{m}\mathbf{a}_i\mathbf{a}_i^T$.  Now, for any $\mathbf{u} \neq \mathbf{0}$, we have  $ \mathbf{u}^T\mathbf{M}^*\mathbf{u}=\sum_{i=1}^{m}(\mathbf{u}^T\mathbf{a}_i)^2>0$. 
%\begin{equation}\label{eq:44}
%  \mathbf{M}^*=\sum_{i=1}^{m}\mathbf{a}_i\mathbf{a}_i^T.
%\end{equation}
%Now, let $\mathbf{u} \neq \mathbf{0}$,
%%\vspace{-.3cm}
%\begin{align}\label{eq:45}
%  0<\mathbf{u}^T\mathbf{M}^*\mathbf{u}=\sum_{i=1}^{m}(\mathbf{u}^T\mathbf{a}_i)^2.
%\end{align}
Thus, $\mathbf{u}\not \in \mathbf{U}^{\perp}$. This implies $\mathbf{U}^{\perp}=\{\mathbf{0}\}$,  and $\mathbf{U}=\left(\mathbf{U}^{\perp}\right)^{\perp}=\mathbb{R}^p \backslash \{\mathbf{0}\}$.
Therefore, if $\mathbf{U}^{\star}=\textup{span}\{\mathbf{a}_t(\boldsymbol\omega),\mathbf{b}_t(\boldsymbol\omega), \mathbf{a}_s(\boldsymbol\omega),\mathbf{b}_s(\boldsymbol\omega):  \boldsymbol\omega \in \mathbb{R}^{p}\}$, then the column space of $\mathbf{M}^*_{FMT}$ defined in   (\ref{eq:43}) is the same as   $\mathbf{U}$,   %$=Span\{\mathbf{a}(\boldsymbol\omega),\mathbf{b}(\boldsymbol\omega);~~ \boldsymbol\omega \in \mathbb{R}^{p}\}$.   %Since matrix $\mathbf{M}^*_{FMT}$ can be written  as matrix $\mathbf{M}$ in Equations \eqref{eq:43} and \eqref{eq:44},  %which it is span of $\mathcal{S}_{\E[y_t\vert \mathbf{Y}_{t-1}]}$ therefore, 
%From   \eqref{eq:43}-\eqref{eq:45}, 
and  $\mathbf{M}^*_{FMT}$ is a real nonnegative definite matrix.   Hence,   for any  $\boldsymbol\beta \in \mathbb{R}^p$, $\boldsymbol\beta^T\mathbf{M}^*_{FMT}\boldsymbol\beta=0$ is equivalent to $\boldsymbol\beta^T\mathbf{a}_j(\boldsymbol\omega)=\boldsymbol\beta^T\mathbf{b}_j(\boldsymbol\omega)=0$ for all $j=t,  s$, and all $\boldsymbol\omega$; which implies that  the column space of $\mathbf{M}^*_{FMT}$, i.e., $\mathcal{S}(\mathbf{M}^*_{FMT})$, is the same as   $\textup{span}\{\mathbf{a}_t(\boldsymbol\omega),\mathbf{b}_t(\boldsymbol\omega), \mathbf{a}_s(\boldsymbol\omega),\mathbf{b}_s(\boldsymbol\omega):\boldsymbol\omega \in \mathbb{R}^p\}$, hence,  from assertion   (a) of  Proposition \ref{prop:2.1},  we have  $\mathcal{S}(\mathbf{M}^*_{FMT})=\mathcal{S}_{\E[y_t\vert \mathbf{Y}_{t-1}]}$.  ~~~~~~~~~~~~~~~~~~~~~~~~~~~~~~~~~~~~~~~~~~~~~~~~~~~~~~~~~~~~~~~~~~~~~~~~~~~~~~~~~~ $\square$

\textbf{Proof of Proposition \ref{prop:2.3}:}  From  Proposition \ref{prop:2.1},  and from \eqref{eq:14}, we have
\begin{equation*}%\label{eq:46}
  \mathbf{M}_{FMT}=\int [\mathbf{a}_t(\boldsymbol\omega) \mathbf{a}_s^T(\boldsymbol\omega)+\mathbf{b}_t(\boldsymbol\omega)\mathbf{b}_s^T(\boldsymbol\omega)]W(\boldsymbol\omega)d\boldsymbol\omega.
\end{equation*}
Since   $W(\boldsymbol\omega)>0$ for $\boldsymbol\omega \in \mathbb{R}^p$,  thus $\mathbf{M}_{FMT}$ is a nonnegative definite matrix, thus  for any $\boldsymbol\beta \in \mathbb{R}^p$ we have,  $ \boldsymbol\beta^T\mathbf{M}_{FMT}\boldsymbol\beta=0 \Longleftrightarrow \boldsymbol\beta^T\mathbf{a}_j(\boldsymbol\omega)=\boldsymbol\beta^T\mathbf{b}_j(\boldsymbol\omega)=0$   for all $j=t,s$,  and all  $\boldsymbol\omega\in \mathbb{R}^p.$
%\begin{align}\label{eq:47}
%  \boldsymbol\beta^T\mathbf{M}_{FMT}\boldsymbol\beta=0 \Longleftrightarrow \boldsymbol\beta^T\mathbf{a}(\boldsymbol\omega)=\boldsymbol\beta^T\mathbf{b}(\boldsymbol\omega)=0 \hspace{.3cm} \text{for all} \hspace{.3cm} \boldsymbol\omega\in \mathbb{R}^p.
%\end{align}
Then, using a similar argument as that in proof of Proposition \ref{prop:2.2}, 
 we have $\mathcal{S}(\mathbf{M}_{FMT})=\textup{span}\{\mathbf{a}_t(\boldsymbol\omega), \mathbf{b}_t(\boldsymbol\omega), \mathbf{a}_s(\boldsymbol\omega), \mathbf{b}_s(\boldsymbol\omega): \boldsymbol\omega \in \mathbb{R}^p\}=\mathcal{S}_{\E[y_t\vert \mathbf{Y}_{t-1}]}$. ~~~~~~~~~~~~~~~~~~~ ~~~~~~~~~~~~~~ ~~~~~~~~~~~~ ~~~~~~~~~~~~~~  $\square$\\
 
%\color{blue}
\textbf{Proof of Theorem \ref{thm:normal}}. %The proof of this theorem can be derive using the results of part (i) in the proof of Lemma 1. 
Since $\widehat{\mathbf{M}}_{FMTn}$ is a V-statistic,  it can  be expressed   as  
\small
\begin{align}\label{eq:normalproof1}
   \widehat{\mathbf{M}}_{FMTn}&=\frac{N-1}{2N}\binom{N}{2}^{-1}\sum_{t<s}\left\{\mathbf{J}_{FMTn}(\mathbf{z}_{t}, \mathbf{z}_{s})
  +\mathbf{J}_{FMTn}^T( \mathbf{z}_{t}, \mathbf{z}_{s})\right\}  
  +\frac{1}{N^2}\sum_{t=p+1}^{N}\mathbf{J}_{FMTn}(\mathbf{z}_{t}, \mathbf{z}_{t}),
 \end{align}
 \normalsize
 where $\mathbf{z}_t= (\mathbf{y}_{t-1},y_t)$ and $\mathbf{z}_s= (\mathbf{y}_{s-1},y_s)$.
The first term on the right side of    \eqref{eq:normalproof1} is a U-statistic,  and the second term is a partial sum of $\mathbf{J}_{FMTn}\left(\mathbf{z}_{t}, \mathbf{z}_{t}\right)$, which is order of $O_p(N^{-1})$.     By  applying   Hoeffding's decomposition and from Lemma \ref{lem:1},   we can write  
\small
\begin{align}
\widehat{\mathbf{M}}_{FMTn}&=\frac{N-1}{2N} \left( 2 \mathbf{M}_{FMTn}+\frac{2}{N}\sum_{t=p+1}^{N}\left(\mathbf{J}_{FMTn}^{(1)}(\mathbf{z}_{t})-2\mathbf{M}_{FMTn}\right)+o_p(N^{-1/2}) \right) + O_p(N^{-1}) \nonumber \\ \label{eq:normalproof2}
&=\mathbf{M}_{FMTn}+\frac{1}{N}\sum_{t=p+1}^{N}\left(\mathbf{J}_{FMTn}^{(1)}(\mathbf{z}_{t})-2\mathbf{M}_{FMTn}\right)+o_p(N^{-1/2}),
\end{align}
 \normalsize
 where $  \mathbf{J}_{FMTn}^{(1)} (\mathbf{z})=\E_{\mathbf{Z}_{s}}\left[\mathbf{J}_{FMTn}\left(\mathbf{z},\mathbf{Z}_{s}\right)+ \mathbf{J}^T_{FMTn}\left(\mathbf{z},\mathbf{Z}_{s}\right) \right]$.  The second term in   \eqref{eq:normalproof2} is the average of a sequence of $m-$dependent  stationary process.  Since  the  covariance matrix of $\vect\left(\mathbf{J}_{FMTn}(\mathbf{Z}_{t},\mathbf{Z}_{s})\right)$ exists,  therefore $\boldsymbol\Sigma_{FMTn}=\Cov\left(\mathbf{J}_{FMTn}^{(1)}(\mathbf{Z}_{t})\right) $  exists,  then by  the central limit theorem for $m-$dependent stationary sequence,  as $N \to \infty$, we have  
%Let $\boldsymbol\Sigma_{FMTn}=\Cov(\mathbf{J}_{FMTn}^{(1)}(\mathbf{y}_{t-1},y_t))$. Then, by the central limit theorem,as $N \to \infty$ we have
\begin{equation*}%\label{eq:normalproof3}
    \sqrt{N}\left(\vect(\widehat{\mathbf{M}}_{FMTn})-\vect(\mathbf{M}_{FMTn})\right) \overset{\mathcal{D}}\longrightarrow \mathcal{N}(\mathbf{0},\boldsymbol\Sigma_{FMTn}). ~~~~~~~~~~~~~~~~~~~~~~~~~~~~~~~~~~~~~~~~~~~~~~~~~~~~~~~~~~~~~~~~~ \square
\end{equation*}

\vspace{.5cm}

\baselineskip=12pt
%\vspace{1cm}
\noindent
{\bf \large References}
\parindent 0pt

\vspace{.1in}

%\bibitem{Bosq}
%D. Bosq, {\em Nonparametric Statistics for Stochastic Processes: Estimation and Prediction}. Lecture Notes in Statistics, Vol. 110, Springer, 1998.
%\bibitem[2]{Cai}
%Z. Cai and E. Masry, {\em Nonparametric estimation of additive nonlinear ARX time series: Local linear fitting and projection}, Vol. 16, No. 4,  Econometric Theory, (2000), pp. 465-501.

%\bibitem[1]{Abraham and Ledolter}
%B. Abraham and J. Ledolter, Statistical Methods for Forecasting, (1983),  Wiley.

%\bibitem[2]{Bass and Clarke}
% F. M. Bass and D. G. Clark,  {\em  Testing distributed lag models of advertising effect}. Journal of Marketing Research, 9, (1972), pp. 298-308.

%\bibitem {CookandLi}
% Cook RD, Li L.(2002). Dimension reduction for the conditional mean in regression{\it Journal of Annals of Statistics}, 30, 455-474.

%\bibitem{ari}
%Aaronson J., Burton R., Dehling H., Gilat D., Hill T., and Weiss B., (1996). Strong laws for L- and U- Statistics, {\em Transactions of the American Mathematical Society}. 348, 2845-2865.

%\bibitem{AIC}
%Akaike, H., (1973). Information Theory and an Extension of the Maximum Likelihood Principle. {\em Proceeding 2nd International Symposium on Informaion Theory, eds. B. N. Petrov and F. Csaki, Budapest: Akademiai Kiado, pp. 267-281}.

%\bibitem{AIC1}
%Akaike, H., (1974). A New Look at the Statistics Model Identification, {\em IEEE Transactions on Automatic Control, AC-19, 716-723.}

%\bibitem{dependetu}
%Bertail, Doukhan, and Soulier (2006), {\em Dependent data in Probability and Statitics}, Springer
\vspace{.07in} \hangindent=15pt    \hangafter=1
%\bibitem{garch}
 Bollerslev T.,  (1986). Generalized Autoregressive Conditional Heteroskedasticity, {\em Journal of Econometrics}. 31, 3, 307-327.

%\bibitem{bro}
%Bradley R. C., (2007). { \em Introduction to Strong Mixing Conditions Volumes 1,2, and 3}. Kendrick Press.

%\bibitem{depedentu}
%Dehlin, Herold (2006), {\em Limit Theorems for Dependent U-Statitics,}
\vspace{.07in} \hangindent=15pt    \hangafter=1
%\bibitem{tst}
Brockwell P.  and Davis  R. (1991).  {\em Time Series: Theory and Methods},  2nd ed. ,  New York: Springer-Verlag.

%\textcolor[rgb]{0.00,0.07,1.00}{
%\bibitem{chester}
%Chet L., (2019). From BigDog to BigDawg: Transitioning an HPC Cluster for Sustainability,{\em Proceedings of Practice and Experience in Advanced Research Computing, PEARC19, July 28-August 1, 2019, Chicago, IL, USA, Association for Computing Machinery (ACM)}.}

%\bibitem{collomb}
%Collomb G., and H$\ddot{a}$rdle W., (1986). Strong uniform convergence rates in robut nonparametric time series analysis and prediction: Kernel regression estimation from dependet observations. {\em Stochastic Processes and their Applications }. 23, 1, 77-89.

\vspace{.07in} \hangindent=15pt    \hangafter=1
%\bibitem{cook94}
Cook R.  D., (1994). On the Interpretation of Regression Plots. {\em Journal of the American Statistical Association}. 89, 425, 177-189.

\vspace{.07in} \hangindent=15pt    \hangafter=1
%\bibitem{cook98}
Cook R.   D., (1998). {\em Regression Graphics: Ideas for Studying Regressions Through Graphics.} {Wiley.}

\vspace{.07in} \hangindent=15pt    \hangafter=1
%\bibitem {coock}
Cook R.  D., (2018).  Principal Components, Sufficient Dimension Reduction, and Envelopes.    {\em Annual Review of Statistics and Its Application}.  5, 533–559.

\vspace{.07in} \hangindent=15pt    \hangafter=1
%\bibitem{cookli}
Cook R.  D., and Li  B., (2002). Dimension Reduction for Conditional Mean in Regression.   {\em   Annals of Statistics}. 30, 455-474.

\vspace{.07in} \hangindent=15pt    \hangafter=1
%\bibitem {cookandweis}
Cook R.  D., and Weisberg S., (1991). Sliced Inverse Regression for Dimension Reduction: Comment. {\em Journal of the American Statistical Association}. 86, 414, 328–332.

%\bibitem {cryer}
%Cryer, J. D., and Chan, K. S. (2008). {\em Time Series Analysis with Applications in R (2nd ed.)}. New York: Springer.

%\bibitem{keller}
%Denker M. and Keller G., (1983). On U-statistics and v. mise’ statistics for weakly dependent processes. {\em Z. Wahrscheinlichkeitstheorie verw Gebiete}. 64, 505–522.

\vspace{.07in} \hangindent=15pt    \hangafter=1
%\bibitem{dependentuv}
Dehling H., (2006).   Limit Theorems for Dependent U-statistics.  In: {\em Dependence in Probability and Statistics.  Lecture Notes in Statistics}. 187, 65-86, New York: Springer. 

\vspace{.07in} \hangindent=15pt    \hangafter=1
%\bibitem{candianlynx}
Elton C., and Nicholson M., (1942). The Ten-Year Cycle in Numbers of the Lynx in Canada. {\em Journal of Animal Ecology.} 11,  2,  215-244.

\vspace{.07in} \hangindent=15pt    \hangafter=1
%\bibitem{engle}
Engle R. F., (1982). Autoregressive Conditional Heteroskedasticity  with Estimates of the Variance of United Kingdom Inflation. {\em Econometrica}. 50, 4, 987-1007.

\vspace{.07in} \hangindent=15pt    \hangafter=1
%\bibitem{Fan:Yao:2003}
Fan J.,  and Yao  Q. (2003).  {\em Nonlinear Time Series: Nonparametric and Parametric Methods},  New York: Springer.

\vspace{.07in} \hangindent=15pt    \hangafter=1
%\bibitem{folland}
Folland G. B., (1992). {\em Fourier Analysis and its Applications.} {Wadsworth and Brooks/Cole Advance Books and Software.}

\vspace{.07in} \hangindent=15pt    \hangafter=1
%\bibitem{Gao:2007}
Gao J.,  (2007).  {\em Nonlinear Time Series: Semiparametric and Nonparametric Methods}.  Chapman and  Hall,  London.

%\bibitem{holder}
%Gilbarg D., and Trudinge N. S., (1983). {\em Elliptic Partial Differential   of Second Order.} {Springer-Verlag New York, Inc}.

\vspace{.07in} \hangindent=15pt    \hangafter=1
%\bibitem{Granger}
Granger  C.  W.  J.,  and Teräsvirta  T.,  (1993).  {\em Modelling Nonlinear Dynamic Relationships},  Oxford. University Press.

 \vspace{.07in} \hangindent=15pt    \hangafter=1
%\bibitem{wolf}
H\"{a}rdle W., Horowitz J.,  and  Kreiss J.,  (2003).   Bootstrap Methods for Time Series. {\em International Statistical Review}. 71, 2, 435-459.

%\bibitem{ADE}
% H\"{a}rdle,  W. , and  Stoker, T.  M.  (1989). Investigating Smooth Multiple Regression by the Method of Average Derivatives. {\em Journal of the American Statistical Association}. 84 (408),  986-995.

%\bibitem{lablaze}
%Honorio J., (2011). Lipschitz Parametrization of Probabilistic Graphical Models. {\em  Proceedings of the Twenty-Seventh Conference on Uncertainty in Artificial Intelligence, 347–354. }

%\bibitem{hris}
%Hristache M., Juditsky A., Polzehl J., and Spokoiny V., (2001). Structure Adaptive Approach for Dimension Reduction. {\em   Annals of Statistics}. 29, 6, 1537-1566.

%\bibitem{tsa}
%Jonathan D. Cryer,\& Kung-Sik Chan, (2008). {\em Time Series Analysis with Applications in R, 2nd Edition}. Springer Science+Business Media, LLC.

\vspace{.07in} \hangindent=15pt    \hangafter=1
%\bibitem{Lehmann}
Lehmann E.L., (1999).  {\em Elements of Large Sample Theory.} {Springer-Verlag New York.}

\vspace{.07in} \hangindent=15pt    \hangafter=1
%\bibitem {li1}
Li K.  C., (1991). Sliced Inverse Regression for Dimension Reduction (with discussion). {\em Journal of the American Statistical Association}. 86, 414, 316-327.

\vspace{.07in} \hangindent=15pt    \hangafter=1
%\bibitem {li2}
Li K.  C., (1992).  On Principal Hessian Directions for Data Visualization and Dimension Reduction: Another Application of Stein's Lemma.  {\em Journal of the American Statistical Association}.  87, 420, 1025-1039.

%\bibitem{cha}
%Li B., Zha H., and Chiaromonte F., (2005). Contour regression: A general approach to dimension reduction. {\em   Annals of Statistics}. 33, 4, 1580-1616.

\vspace{.07in} \hangindent=15pt    \hangafter=1
%\bibitem{Ma:Zhu:2013}
Ma  Y.,  and Zhu  L.  P.,  (2013).  Efficient Estimation in Sufficient Dimension Reduction.  {\em   Annals of Statistics},  41, 1,  250-268.

\vspace{.07in} \hangindent=15pt    \hangafter=1
%\bibitem{Masry:Fan:1997}
Masry E., and Fan  J.,  (1997).  Local Polynomial Estimation of Regression Functions for Mixing Processes.   {\em Scandinavian Journal of Statistics},  24,  2,  165–179.

\vspace{.07in} \hangindent=15pt    \hangafter=1
%\bibitem{moran}
Moran P. A. P., (1953). The Statistical Analysis of the Canadian Lynx Cycle. {\em Australian Journal of Zoology}. 1, 3, 291–298.

%XSEDE site
%\textcolor[rgb]{0.00,0.07,1.00}{
%\bibitem{campuschamp}
%Brazil M., Brunson D., Culich A., DeStefano L., Jennewein D., Jolley T., Middelkoop T., Neeman H., Rivera L., Smith J., and Wernert J., (2019). Campus Champions: Building and sustaining a thriving community of practice around research computing and data Proceedings of Practice and Experience in Advanced Research Computing. {\em PEARC19, Chicago, IL, USA, Association for Computing Machinery (ACM)}}.
%
%\textcolor[rgb]{0.00,0.07,1.00}{
%\bibitem {Nystrom}
%Nystrom N. A., Levine M. J., Roskies R. Z., and Scott J. R., (2015). Bridges: A Uniquely Flexible HPC Resource for New Communities and Data Analytics. In Proceedings of the 2015 Annual Conference on Extreme Science and Engineering Discovery Environment. {\em St. Louis, MO, July 26-30, 2015. XSEDE15. ACM, New York, NY, USA.}}

\vspace{.07in} \hangindent=15pt    \hangafter=1
%\bibitem {parksamadi1}
 Park J.  H., and Samadi S. Y., (2014). Heteroscedastic Modelling via the Autoregressive Conditional Variance Subspace. {\em The Canadian Journal of Statistics}. 42, 3, 423-435.

\vspace{.07in} \hangindent=15pt    \hangafter=1
% \bibitem {parksamadi2}
 Park J.  H., and Samadi S. Y., (2020). Dimension Reduction for the Conditional Mean and Variance Functions in Time Series. {\em Scandinavian Journal of Statistics}. 47, 1, 134-155.

\vspace{.07in} \hangindent=15pt    \hangafter=1
%\bibitem {tscms}
Park J.  H., Sriram T.  N., and  Yin X., (2009).  Central Mean Subspace in Time Series. {\em Journal of Computational and Graphical Statistics}. 18, 3, 717-730.

\vspace{.07in} \hangindent=15pt    \hangafter=1
%\bibitem {tscs}
Park J.  H., Sriram T.  N., and  Yin X.,(2010). Dimension Reduction in Time Series. {\em Statistica Sinica}. 20, 747-770.

%\bibitem {DongWang}
%Dong Wang, Haipeng Shen, \&  Young Truong (2015).  Efficient dimension reduction for high-dimensional matrix valued data  {\it Journal of Neurocomputing}.

\vspace{.07in} \hangindent=15pt    \hangafter=1
%\bibitem{scotte}
Scott D. W., (1992). {\em Multivariate Density Estimation: Theory, Practice, and Visualization.} { John Wiley and Sons, Inc.}

\vspace{.07in} \hangindent=15pt    \hangafter=1
%\bibitem{silverman}
Silverman B.  W., (1986). {\em Density Estimation for Statistics and Data Analysis}. Chapman and Hall, New York.

%\bibitem{lip}
%Sohrab H. H., (2003). {\em Basic Real Analysis} . Birkhäuser.

\vspace{.07in} \hangindent=15pt    \hangafter=1
%\bibitem{Tiao:Tsay:1994}
Tiao G. C.,  and Tsay  R. S. (1994).  Some Advances in Nonlinear and Adaptive Modeling in Time Series.  {\em Journal of Forecasting}. 13, 2,  109-131.

\vspace{.07in} \hangindent=15pt    \hangafter=1
%\bibitem{tong:1990}
Tong H., (1990). {\em Nonlinear Time Series: A Dynamical Systems Approach.} { Oxford University Press (UK).}

\vspace{.07in} \hangindent=15pt    \hangafter=1
%\bibitem{tong:1995}
Tong H., (1995).   A Personal Overview of Nonlinear Time Series Analysis from a Chaos Perspective (with discussion).  {\em Scandinavian Journal of Statistics}.  22, 4,  399–445.

%\bibitem {Bowman}
%Bowman, A.~W., \& Azzalini, A. (1997).  {\it Applied Smoothing Techniques for Data Analysis},  London, Oxford University Press.

\vspace{.07in} \hangindent=15pt    \hangafter=1
%\bibitem{tstay}
Tsay R.  S., (1988). Non-Linear Time Series Analysis of Blowfly Population.  {\em Journal of Time Series Analysis}.  9, 3,247-263.
%Tsay R.S., (1988). Non-Linear Time Series Analysis of Blowﬂy Population.  {\em Journal of Time Series Analysis}.  9, 3, 247-263.
%xsede site

%
%\textcolor[rgb]{0.00,0.07,1.00}{
%\bibitem{Towns}
%Towns J., Cockerill T., Dahan, M., Foster I., Gaither K., Grimsha, A., Hazlewood, V., Lathrop S., Lifka D., Peterson G.D., Roskies R., Scott, J.R. and Wilkens-Diehr N., (2014). XSEDE: Accelerating Scientific Discovery. Computing in Science \& Engineering. {\em IEEE Computer Society.} 16, 5, 62-74. }

\vspace{.07in} \hangindent=15pt    \hangafter=1
%\bibitem{Tjostheim}
Tj{\o}stheim   D.  (1994).  Nonlinear Time Series: A Selective Review,  {\em Scandinavian Journal of Statistics}.  21, 2,  97–130.

\vspace{.07in} \hangindent=15pt    \hangafter=1
%\color{blue} we should cite this one ! \color{black}
%\bibitem{fmtsnew}
Weng J., and Yin X., (2018). Fourier Transform Approach for Inverse Dimension Reduction Method. {\em Journal of Nonparametric Statistics}. 30, 4, 1029-0311.

\vspace{.07in} \hangindent=15pt    \hangafter=1
%\bibitem{an}
Xia Y. , and An H. Z., (1999). Projection Pursuit Autoregression in Time Series. {\em Journal of Time Series Analysis}. 20, 6, 693–714.

\vspace{.07in} \hangindent=15pt    \hangafter=1
%\bibitem{Xia:Li:1999}
Xia  Y., and Li  W.  K. (1999).  On Single-Index Coefficient Regression Models.  {\em Journal of the American Statistical Association}.   94,  448,   1275-1285.

\vspace{.07in} \hangindent=15pt    \hangafter=1
%\bibitem{Xia:Tong:Li:1999}
Xia Y., Tong  H. and Li  W. K. (1999).  On Extended Partially Linear Single-Index Models. {\em Biometrika}.   86,  831-842.

\vspace{.07in} \hangindent=15pt    \hangafter=1
%\bibitem{xiaatel}
Xia Y., Tong H., Li W. K., and Zhu L. X., (2002). An Adaptive Estimation of Dimension Reduction. {\em Journal of the Royal Statistical Society. Series B}. 64, 3, 363–410.

\vspace{.07in} \hangindent=15pt    \hangafter=1
%\bibitem{yeandwesi}
Ye Z., and Weiss  R.  E., (2003). Using the Bootstrap to Select One of a New Class of Dimension Reduction Methods.  {\em Journal of the American Statistical Association}. 98, 464, 968-979.

\vspace{.07in} \hangindent=15pt    \hangafter=1
%\bibitem{yincook02}
Yin X., and Cook R. D., (2002). Dimension Reduction for the Conditional kth Moment in Regression. {\em Journal of the Royal Statistical Society: Series B}.  64, 2, 159–175.

\vspace{.07in} \hangindent=15pt    \hangafter=1
%\bibitem {yinandli}
Yin X., and Cook R. D., (2005). Direction estimation in single-index regression. {\em Biometrika}. 92, 2, 371–384.

\vspace{.07in} \hangindent=15pt    \hangafter=1
%\bibitem {yincook}
Yin X., Li B., and Cook R. D., (2008). Successive direction extraction for estimating the central subspace in a multiple-index regression. {\em Journal of Multivariate Analysis}. 99, 8, 1733-1757.

%\bibitem {yoshi}
%Yoshihara K., (1976). Limiting behaviour of U-Statistics for stationary, absolutely regular process. {\em Zeitschrift f$\ddot{u}$r Wahrscheinlichkeitstheorie und Verwandte Gebiete}. 35, 237-252.

%\bibitem{Davydov}
% Davydov, Yu.~A. (1973).  Mixing conditions for Markov chains,  {\it Theory of Probability and Its Applications}, 18, 312-328.
%\bibitem{Zeng and Zhu.a}

\vspace{.07in} \hangindent=15pt    \hangafter=1
Zeng P.,  and  Zhu Y.,  (2010). An Integral Transform Method for Estimating the Central Mean and Central Subspaces. {\em Journal of Multivariate Analysis}. 101, 1, 271-290.

\vspace{.07in} \hangindent=15pt    \hangafter=1
%\bibitem{Zeng and Zhu.b }
Zhu Y., and Zeng P., (2006). Fourier Methods for Estimating the Central Subspace and the Central Mean Subspace in Regression  {\em Journal of the American Statistical Association}. 101, 476, 1638-1651.

\vspace{.07in} \hangindent=15pt    \hangafter=1
%\bibitem{Zhu and Zhu:2009}
Zhu  L.  P.,  and  Zhu  L.  X.,  (2009).  Dimension Reduction for Conditional Variance in Regression.  {\em Statistica Sinica},   19,  869-883.
%\end{thebibliography}

%\newpage
% \section*{Tables and Figures}

%\vspace{.2cm}

%%%%%%%%%%%%%%%%%%%%%%

\end{document}